\documentclass[11pt]{article}
\usepackage[utf8]{inputenc}
\pdfoutput=1
\usepackage{amsmath,amssymb,graphicx,multirow,xspace}
\usepackage[colorlinks=true,urlcolor=blue,anchorcolor=blue,citecolor=blue,filecolor=blue,linkcolor=blue,menucolor=blue,linktocpage=true]{hyperref}
\usepackage[compress,numbers]{natbib}
\usepackage{braket}
\usepackage{slashed}
\usepackage{physics}
\usepackage{bm}
\usepackage{indentfirst}
\usepackage{appendix}
\usepackage{verbatim}
\usepackage{cancel}
\usepackage{cleveref}
\crefname{equation}{Eq.}{Eqs.} 
\crefrangelabelformat{equation}{(#3#1#4--#5#2#6)}

\usepackage[T1]{fontenc}
\long\def\symbolfootnote[#1]#2{\begingroup%
  \def\thefootnote{\fnsymbol{footnote}}\footnote[#1]{#2}\endgroup}

\newcommand{\gev}{\mathrm{GeV}}
\newcommand{\tev}{\mathrm{TeV}}

\newcommand{\hvev}[1]{ \left\langle {#1} \right\rangle }

\input prepictex
\input pictex
\input postpictex
\newdimen\tdim
\tdim=\unitlength

\begin{document}
\begin{titlepage}

    \vspace{0.5cm}
    \begin{center}
        \Large\bf
        ALP-Assisted Strong First-Order Electroweak \\ Phase Transition and Baryogenesis
    \end{center}

    \vspace{0.2cm}
    \begin{center}
        {
            Keisuke Harigaya$^{1,2,3}$\symbolfootnote[1]{kharigaya@uchicago.edu}
            and Isaac R. Wang$^4$\symbolfootnote[2]{isaac.wang@rutgers.edu}
        }\\
        \vspace{0.6cm}
        \textit{
            \vspace{0.5cm}
            $\,^1$ Department of Physics, University of Chicago, Chicago, IL 60637, USA \\
            \vspace{0.5cm}
            $\,^2$ Enrico Fermi Institute and Kavli Institute for Cosmological Physics, University of Chicago, Chicago, IL 60637, USA \\
            \vspace{0.5cm}
            $\,^3$ Kavli Institute for the Physics and Mathematics of the Universe (WPI),\\
            The University of Tokyo Institutes for Advanced Study,\\
            The University of Tokyo, Kashiwa, Chiba 277-8583, Japan \\
            \vspace{0.5cm}
            $\,^4$New High Energy Theory Center,
            Department of Physics and Astronomy,\\
            Rutgers University, Piscataway, NJ 08854, USA}\\
    \end{center}

    \vspace{0.4cm}

    \begin{abstract}
        Axion-like particles (ALPs) can be naturally lighter than the electroweak scale. We consider an ALP that couples to the Standard Model Higgs to achieve the strong first-order electroweak phase transition. We discuss the two-field dynamics of the phase transition and the associated computation in detail and identify the viable parameter space. The ALP mass can be from the MeV to GeV scale.
        Baryon asymmetry can be explained by local baryogenesis without violating the current electron and atom electric dipole moment bound in most of the viable parameter space.
        The viable parameter space can be probed through Higgs exotic decay, rare kaon decay, the electron and atomic electric dipole moment, and the effective number of neutrinos in the cosmic microwave background in the future.
        The gravitational-wave signal is too weak to be detected.
    \end{abstract}

\end{titlepage}

\vspace{0.2cm}
\noindent

\tableofcontents
\newpage
\section{Introduction}

The predominance of matter over antimatter in the universe is a well-established fact. However, despite the success of the Standard Model (SM) of particle physics, the origin of matter-antimatter asymmetry, referred to as baryogenesis, is still obscure.
As pointed out by Sakharov,  successful baryogenesis mechanisms must contain baryon number (B) violation, C and CP violation, and out-of-thermal-equilibrium processes~\cite{Sakharov:1967dj}.
Since the first proposal in 1985~\cite{Kuzmin:1985mm}, electroweak baryogenesis (EWBG)
has drawn significant attention.
Indeed, the baryon symmetry is anomalous at the quantum level in the SM and is violated by the weak sphaleron process~\cite{Kuzmin:1985mm,tHooft:1976rip,tHooft:1976snw,Manton:1983nd,Klinkhamer:1984di}.
Moreover, a CP-violating phase exists in the CKM matrix of the weak sector~\cite{Cabibbo:1963yz,Kobayashi:1973fv}.
If the electroweak phase transition (EWPT) is a strong first-order phase transition (SFOPT), the out-of-equilibrium condition will also be satisfied.
Nevertheless, the SM EWPT has been shown to be a smooth crossover that cannot satisfy the out-of-equilibrium condition~\cite{Kajantie:1993ag,Farakos:1994kj,Jansen:1995yg,Kajantie:1995kf,Rummukainen:1996sx,Kajantie:1996mn,Gurtler:1997hr,Csikor:1998eu,Laine:1998vn,Laine:1998qk,Rummukainen:1998as,Fodor:1999at}.
Furthermore, even if the phase transition is a SFOPT, the resulting baryon asymmetry from the EWBG in the minimal SM has been found to be much smaller than the observed value due to the tiny Yukawa couplings of quarks and the CKM mixing~\cite{Gavela:1993ts,Huet:1994jb,Gavela:1994ds,Gavela:1994dt,Gavela:1994yf}.
Therefore, new physics is necessary in order to enhance the strength of the EWPT and provide additional CP-violation.
One way to enhance the strength of the phase transition is to extend the SM scalar sector with an extra singlet scalar that generates trilinear terms at tree level~\cite{Pietroni:1992in,Choi:1993cv,Ham:2004cf,Noble:2007kk,Ahriche:2007jp,Profumo:2007wc,Barger:2007im,Barger2008,Das:2009ue,Ashoorioon:2009nf,Barger:2011vm,Espinosa:2011ax,Chung:2012vg,Patel:2012pi,Huang:2014ifa,Curtin:2014jma,Jiang:2015cwa,Kotwal:2016tex,Tenkanen:2016idg,Chen:2017qcz,Chiang:2017nmu,Kurup:2017dzf,Carena:2019une,Friedlander:2020tnq,Azatov:2022tii}. Such trilinear terms generate a barrier at finite temperature between the true vacuum and the false vacuum during the phase transition and thus can enhance the strength of the EWPT. These models often predict signals of extra scalar production, Higgs exotic decay, and gravitational waves (GWs). Moreover, additional CP-violation may exist in the scalar sector, which typically can be observed in future electron electric dipole moment (EDM) measurements.

To affect the EWPT, the scalar particle must be lighter or at most around the electroweak scale. This leads to a question concerning naturalness akin to that of Higgs: Why can this singlet have a mass much smaller than other fundamental scales? In addition, in light of the current extensive and comprehensive experimental search for dark-scalar mixing with the Higgs, the scalars are required to be weakly interacting with the Higgs, and, naturally, with other SM particles. These two hints motivate considering scalar particles that are naturally light and weakly interacting.

One well-known example of this type of particles is the axion-like particle (ALP).  The ALP is the angular degree of freedom of a complex scalar $P = |P| \exp (i S/f)$ with rotational $U(1)$ symmetry. This symmetry is spontaneously broken at the energy scale $f$, leaving $S$ as the remaining singlet scalar particle at the EW scale (or below), with a shift symmetry $S \rightarrow S+\delta S$. The shift symmetry can be explicitly broken to give a potential of $S$ and coupling with the Higgs.
The fact that the ALP is the pseudo-Nambu-Goldstone boson (pNGB) makes it naturally light and weakly interacting.

The potential of $S$ and its leading order coupling with the Higgs is~\cite{Jeong:2018ucz}
\begin{align}
    \label{eq:full model}
    V(H,S) = - (\mu_H^2  - A f \cos \delta)|H|^2 + \lambda |H|^4 + \mu_S^2 f^2 \left(1 - \cos (\frac{S}{f})\right) - A f \left(|H|^2 - v^2\right) \cos (\frac{S}{f} - \delta),
\end{align}
where $v \simeq 174~\gev$ is the measured EW vacuum expectation value (vev) of the Higgs field.
Here we choose the minima of $S$ after the EWPT to be $S = 0 + 2n\pi$ by the shift of $S$.
There is a phase difference $\delta$ between the potential of $S$ and the interaction term with the Higgs.
This phase can arise from CP-violation in the UV completion of the model.

In large $f$ limit, the potential becomes
\begin{align}
    \label{eq:original}
    V(H,S) = - \mu_H^2 |H|^2 + \lambda |H|^4 + \frac{1}{2}\mu_S^2 S^2 - A' S (|H|^2 - v^2),
\end{align}
where $A' \equiv A \sin \delta$.
This potential was first introduced in~\cite{Das:2009ue} and reviewed in~\cite{Espinosa:2011ax}.
The naturalness of the lightness of $S$ is pointed out in~\cite{Harigaya:2022ptp}, with a thorough investigation of the two-field dynamics of the phase transition, vacuum metastability, and MeV scale parameter space.
We call the model with the potential in  Eq.~\eqref{eq:original} ``the simplified model'' throughout this paper.
This potential, along the path $S(H)$ where $\partial V/\partial S = 0$, becomes
\begin{align}
    \label{eq:origianl 1d}
    V(H,S(H)) = - \frac{1}{2}\mu_H^2 h^2 + \frac{1}{4}(\lambda - \frac{A'^2}{2 \mu_S^2})h^4.
\end{align}
The effective quartic coupling, $\lambda - A'^2/(2\mu_S^2)$, becomes small and this is often presumed to make the EWPT strongly first-order because the thermal effect becomes more important in comparison with the zero-temperature potential. Here the strong first-order EWPT is usually defined to be $v_c/T_c \geq 1$, where $T_c$ denotes the critical temperature of the phase transition and $v_c$ is the Higgs vev at $T_c$.

This conclusion holds true if the finite-temperature effective potential is computed under high-temperature expansion without proper resummation and one-loop zero-temperature correction
(for example, see~\cite{Quiros:1999jp}.)
Such a computation technique is also used in previous works for the simplified model~\cite{Das:2009ue,Espinosa:2011ax,Harigaya:2022ptp}. However, in the SM with a small quartic coupling (i.e., a small Higgs mass), once thermal resummation and the Coleman-Weinberg correction are properly included, the EWPT is significantly weakened and is not strong  enough~\cite{Arnold:1992rz} (see~\cite{Lofgren:2023sep} for a recent review.).
The main suppression comes from the top quark Yukawa.
We performed a two-loop computation with the state-of-the-art dimensional reduction method with the help of the \verb|DRalgo| package~\cite{Ekstedt:2022bff} to check the viability of the simplified model. (See~\cite{Ginsparg:1980ef,Appelquist:1981vg,Braaten:1995cm,Kajantie:1995dw} for the original references of dimensional reduction and~\cite{Croon:2020cgk,Gould:2021dzl,Niemi:2021qvp,Schicho:2021gca} for recent applications and reviews.)
We find that the EWPT is not strong enough to prevent the washout of baryon number for the simplified model~\cite{Das:2009ue,Espinosa:2011ax,Harigaya:2022ptp}, see Appendix~\ref{sec:app}, where we show the numerical results of various computation methods and comment on the uncertainties.
Although higher-order computations and/or more careful treatment on various sources of uncertainties may finally reveal that the simplified model can work,
the viability of the simplified model is questionable.%
\footnote{Previous lattice simulations claiming an SFOPT with a small Higgs mass $m_H \simeq 35~\gev$ in the SM were based on pure $SU(2)+\text{Higgs}$ theory, and thus the negative contribution from top Yukawa was not included. For example, see Ref.~\cite{Kajantie:1993ag,Jansen:1995yg,Kajantie:1995kf,Kajantie:1996mn,Fodor:1999at}. Those full SM simulations with a proper top quark mass considered only reached a physical Higgs mass $m_H \simeq 50~\gev$, corresponding to $v_c/T_c \simeq 0.7$, which is consistent with the two-loop computation of Ref.~\cite{Arnold:1992rz}.}.

In this work, we consider the full potential of an ALP in Eq.~\eqref{eq:full model}, emphasizing the deviation from the simplified model as $f$ gets smaller. With a smaller UV scale $f$, the higher-order terms in $S$ may be effective and enhance the EWPT. This work focuses on this full potential, with the thermal effective potential computed at one-loop level with proper resummation and zero-temperature corrections. We show that the full ALP+SM model does give an SFOPT in a wide range of the parameter space. This enhancement does not require a large mixing angle between the ALP and Higgs, and the model works even for a MeV scale ALP, and thus opens a window for a SFOPT by extra scalars.

In addition to the enhancement of the phase transition strength, this model provides chances to produce the observed amount of baryon asymmetry in the universe (BAU) via the local electroweak baryogenesis mechanism~\cite{Dine:1990fj,Dine:1992ad,Trodden:1998ym}. Traditional local electroweak baryogenesis considers an effective operator $|H|^2 W \tilde{W}/M^2$, where $M$ is some UV scale, to produce a baryon number asymmetry proportional to $(\Delta h/M)^2$. Since $\delta h$ is around the weak scale, to explain the observed BAU requires the UV scale $M$ not much above the weak scale, which leads to a too large electron EDM. The mechanism was ruled out shortly after its proposal. In the model we consider, on the other hand, we may consider the dimension-5 couplings of the ALP with the $W$ boson or the $SU(2)$ charged fermions suppressed by a UV scale $M$.
Because the shift of $S$ during the phase transition is much above the weak scale, the UV scale $M$ may be much above the weak scale,  so the EDM constraint is avoided in most of the viable parameter spaces.
Because the shift of $S$ during the phase transition is much above the weak scale, the UV scale $M$ may be much above the weak scale,  so the EDM constraint is avoided in most of the viable parameter spaces.

We discuss various ways to probe this ALP model.
Unfortunately, we find that the gravitational-wave signal from the EWPT is too weak to be detected. But scalar direct production can put a stringent constraint on the GeV scale ALP, which can be probed in the future by Higgs exotic decay. The MeV scale ALP can be probed by beam-dump experiments searching for rare Kaon decays, such as NA62 or KLEVER.
Furthermore, the MeV scale ALP contributes negatively to the effective neutrino numbers of the universe, which can be probed by the future Cosmic Microwave Background (CMB)-S4 experiment.

The coupling of the Higgs with a NGB to affect the EWPT was discussed in the literature~\cite{Espinosa:2011eu,Jeong:2018jqe,Jeong:2018ucz,Bian:2019kmg,DeCurtis:2019rxl}. In~\cite{Espinosa:2011eu,Bian:2019kmg,DeCurtis:2019rxl}, the Higgs is a composite field to solve the electroweak hierarchy problem and is also a NGB. In this paper, we consider the case where the Higgs is a fundamental scalar field and do not address the electroweak hierarchy problem by the compositeness of the Higgs. The smallness of the electroweak scale may be explained by anthropic requirements~\cite{Agrawal:1997gf,Hall:2014dfa,DAmico:2019hih} or (partially) by supersymmetry~\cite{Maiani:1979cx,Veltman:1980mj,Witten:1981nf,Kaul:1981wp}.
Refs.~\cite{Jeong:2018ucz,Jeong:2018jqe} consider the same IR model in Eq.~\eqref{eq:original} as well as local baryogenesis by the coupling of $S$ with the $SU(2)_L$ gauge bosons. We include one-loop corrections to the zero-temperature potential as well as resummation in the computation of the thermal potential, which can significantly suppress the strength of the PT as described above.
The one-loop corrections to the zero-temperature potential can also destabilize the electroweak vacuum, but we will show that the lifetime of the vacuum is still long enough.
We also present different UV completions from Refs.~\cite{Jeong:2018ucz,Jeong:2018jqe}.

This paper is organized as follows. In Sec.~\ref{sec:model}, we discuss the basic settings of the model as well as the loop corrections. Naturalness requirement and metastability are also discussed in this section. In Sec.~\ref{sec:phase transition}, we discuss the thermal phase transition in detail. In Sec.~\ref{sec:ewbg}, we discuss how EWBG can be achieved in this model via the local baryogenesis mechanism.
A quick review of local electroweak baryogenesis is presented.
Sec.~\ref{sec:UV} provides UV completion. In Sec.~\ref{sec:probe}, we discuss how to probe this model in various experiments and astrophysical observations. Finally, we summarize this work in Sec.~\ref{sec:conclusion}.

\section{The model}
\label{sec:model}

\subsection{The tree-level potential}
\label{sec:tree-level}
The Higgs doublet is decomposed as
\begin{align}
    \label{eq:Higgs component}
    H = \frac{1}{\sqrt{2}} \begin{pmatrix}
                               \chi_1 + i \chi_2 \\
                               h + i \chi_3
                           \end{pmatrix},
\end{align}
where $h$ is the physical Higgs field that obtains the vev $\langle h \rangle = \sqrt{2} v$ and $\chi_{1,2,3}$ are the would-be Nambu-Goldstone modes. The ALP-Higgs scalar potential in Eq.~\eqref{eq:full model} is written as
\begin{align}
    \label{eq:full model physical}
    V = - \frac{1}{2} (\mu_H^2 - A f \cos \delta) h^2 + \frac{1}{4}\lambda h^4 + \mu_S^2 f^2 \left(1 - \cos (\frac{S}{f})\right) - \frac{1}{2}A f \left(h^2  -2v^2\right) \cos (\frac{S}{f} - \delta).
\end{align}
The periodicity in $S$ of the third and fourth terms can be in general different from each other, but we consider the case with the same periodicity. The UV completion presented in Sec.~\ref{sec:UV} indeed gives the same periodicity.
We take $\delta \in [-\pi,\pi]$ without loss of generality.
Noticing the spurious symmetry $S \rightarrow -S$ and $\delta \rightarrow - \delta$ in the potential, we may further take $\delta \in [0, \pi]$.
For $\delta > \pi/2$, we find that the electroweak vacuum with $\langle h \rangle \simeq \sqrt{2}v$ is not the absolute minimum, and there exists a deeper minimum with larger $\langle h \rangle$. We may redefine the parameters so that at that minimum $\langle h \rangle = \sqrt{2} v$, and the corresponding $\delta $ after the shift of $S$ to set $S=0$ becomes smaller than $ \pi/2$.
In conclusion, we only concentrate on $0 \leq \delta \leq \pi/2$.

The would-be Nambu-Goldstone boson masses and the physical scalar mass matrix are
\begin{align}
    \label{eq:scalar mass matrix}
    m_{\rm gs}^2 & = \lambda h^2 - \mu_H^2 - A f \cos \left( \frac{S}{f} - \delta \right) + A f \cos \delta, \nonumber                                                                                             \\
    m_{h,S}^2    & = \begin{pmatrix}m^2_{hh} & m^2_{hS} \\ m^2_{hS} & m^2_{SS}\end{pmatrix} \nonumber                                                                                                              \\
                 & \equiv  \begin{pmatrix}
                               3\lambda h^2 - \mu_H^2 - A f \cos \left( \frac{S}{f} - \delta \right) + A f \cos \delta & A h \sin \left( \frac{S}{f} - \delta \right)                                                  \\
                               A h \sin \left( \frac{S}{f} - \delta \right)                                            & \mu_S^2 \cos \frac{S}{f} + \frac{A}{2f} (h^2 - 2v^2) \cos \left( \frac{S}{f} - \delta \right)
                           \end{pmatrix}.
\end{align}
The $h$ and $S$ boson mass squares are the eigenvalues of the mass matrix,
\begin{align}
    \label{eq:scalar eigenvalue}
    m^2_{\pm} = \frac{1}{2} \left( m^2_{hh} + m^2_{SS} \pm \sqrt{(m^2_{hh} - m^2_{SS})^2 + 4 m^4_{hS}} \right).
\end{align}
We consider the case where the heavier mass eigenstate with a mass square $m_+^2$ is the SM-like Higgs, and the lighter one is a new state.

\begin{figure}[!t]
    \centering
    \includegraphics[width=0.45\linewidth]{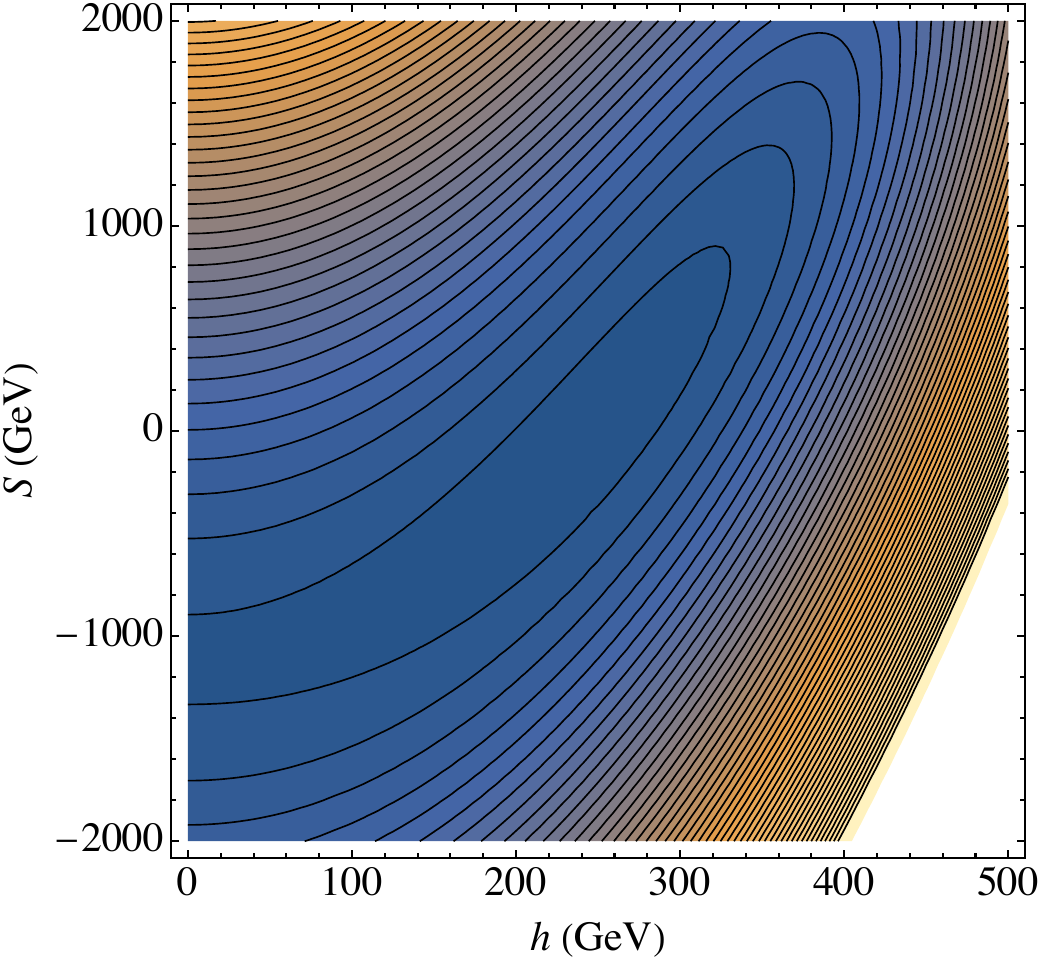}
    \includegraphics[width=0.452\linewidth]{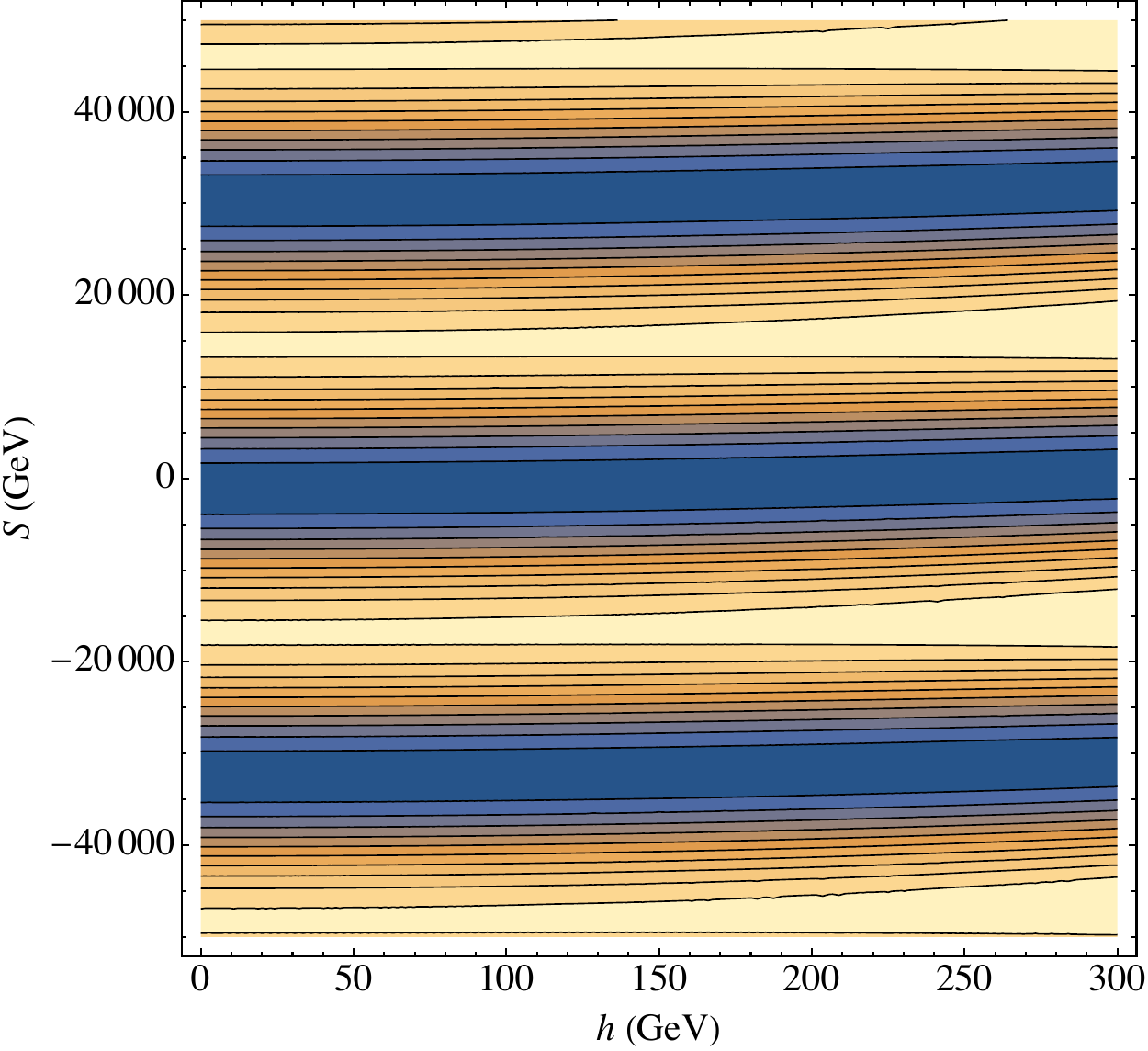}
    \caption{Tree-level potential for a benchmark point with $f = 5~\tev$, $\delta = 2\pi/5$, $m_S = 5~\gev$, and $\sin \theta = 0.1$. Left: potential in a small range of $S$. One can clearly see the ``valley" that passes through the point $(\sqrt{2}v, 0)$. Right: potential in a wider range of $S$. One can see the periodicity of the potential along the $S$ direction.}
    \label{fig:tree-potential}
\end{figure}

The free parameters in this Lagrangian are $\{\lambda, \mu_H, \mu_S, A, f, \delta\}$.
While keeping $\delta$ and $f$ as free parameters, we express the remaining four by other quantities that are directly observable: the Higgs vev $v \simeq 174~\gev$, the Higgs mass $m_h \simeq 125~\gev$, the ALP mass $m_S$, and the mixing angle between the ALP and Higgs $\theta$.
At the tree level, the relation is given by
\begin{align}
    \label{eq:tree-level param}
    A       & = \frac{1}{2\sqrt{2}v}(m_h^2 - m_S^2)\sin(2\theta)\csc(\delta),\nonumber            \\
    \lambda & = \frac{1}{8v^2}\left(m_h^2 + m_S^2 + (m_h^2 - m_S^2)\cos(2\theta)\right),\nonumber \\
    \mu_H^2 & = \frac{1}{4}\left(m_h^2 + m_S^2 + (m_h^2 - m_S^2)\cos(2\theta)\right),\nonumber    \\
    \mu_S^2 & = \frac{1}{2} \left(m_h^2 + m_S^2 - (m_h^2 - m_S^2)\cos(2\theta)\right).
\end{align}
Matching $A \sin \delta$ in this model with $A'$ in the simplified model, Eq.~\eqref{eq:tree-level param} is exactly the same as the parametrization in the simplified model~\cite{Das:2009ue,Harigaya:2022ptp}.
For example, at a benchmark point $f = 5~\tev$, $\delta = 2\pi/5$, $m_S = 5~\gev$, and $\sin \theta = 0.1$, Eq.~\eqref{eq:tree-level param} gives
\begin{align}
    \label{eq:tree-level benchmark}
    A = 6.65~\gev,~\mu_S^2 = 181.325~\gev^2,~\mu_H^2 = 7750.6 ~\gev^2,~\lambda = 0.128.
\end{align}
The potential shape at this benchmark point is plotted in Fig.~\ref{fig:tree-potential}.

To get the path that minimizes the potential, i.e., the ``valley'' of the potential in the $h-S$ space, we take $\partial V/\partial S = 0$. The field value of S along this path can be written as
\begin{align}
    S (h)= f \arctan \left(\frac{A (h^2 - 2v^2) \sin (\delta)}{2f \mu_S^2 + A (h^2 - 2v^2) \cos (\delta)} \right).
\end{align}
The field value shift of $S$ when $h$ shift from $0$ to $\sqrt{2}v$ is thus roughly
\begin{align}
    \label{eq:Spath}
    \Delta S \simeq \frac{A}{\mu_S^2}v^2 \sin \delta.
\end{align}
where the large $f$ limit was taken.

When $\Delta S \ll f$, the dynamics of the model can be well-described by the simplified model, while as $\Delta S$ gets closer to $f$, the deviation from the simplified model becomes significant and the phase transition may be affected.  We thus define the critical decay constant $f_c$ via
\begin{align}
    \label{eq:fc def}
    f_c \equiv \frac{A}{\mu_S^2} v^2 \sin \delta,
\end{align}
and parameterize $f$ as $f \equiv c f_c$. When scanning over the parameter space, we use $c$ instead of $f$ as a free parameter.
$f_c$ is defined by the tree-level parameters. The one-loop parameters are then solved from this $f_c$, following the framework discussed in the next subsection.

We require $V(\sqrt{2}v, S (\sqrt{2}v)) < V(0, S (0))$ for the stability of the EW vacuum. The precise expression is complicated, but it roughly behaves as $\mu_S^2 f^2 > A v^2 f$ in large $f$ limit.
Expressing $f$ by $c$, the bound is
\begin{align}
    \label{eq:ew-stability}
    c \sin \delta > 1.
\end{align}
This lower bound on $c$ strongly constrains the model.
Numerical computation indeed shows that $c$ slightly below this value will lead to the violation of the stability requirement for almost all the parameter space.
As will be discussed in Sec.~\ref{sec:phase transition},  $c$ needs to be as small as possible to achieve a stronger phase transition.
In the parameter space with small $\delta$, increasing $c$ is necessary. This will be discussed in detail in Sec.~\ref{sec:phase transition}.
In addition, since the precise expression of the condition is not exactly Eq.~\eqref{eq:ew-stability}, especially when 1-loop quantum correction is taken into account, we numerically check the stability requirement during parameter scanning, and show the excluded parameter space in gray shaded region in Fig.~\ref{fig:final result} and~\ref{fig:edm-final}.

For small $c$, the UV cutoff of the theory is strongly constrained by naturalness.
The quantum correction to the potential of $S$ by the Higgs loop is
\begin{align}
    \label{eq:close loop}
    \Delta \mu_S^2 \sim \frac{\Lambda_H^2}{16 \pi^2} \frac{A}{f},
\end{align}
where $\Lambda_H$ is the cutoff of the loop integral.
We require that $\Delta \mu_S^2 < \mu_S^2$. Using $f = c f_c$ and Eq.~\eqref{eq:fc def}, this condition becomes
\begin{align}
    \label{eq:naturalness final}
    \Lambda_H < 4\pi v  \sqrt{c \sin \delta}.
\end{align}
For $c\sin \delta= O(1)$, the cutoff should be around the TeV scale.
This does not necessarily require the significant modification of the Higgs itself around the TeV scale by such as supersymmetry or compositeness of the Higgs. In Sec.~\ref{sec:UV}, we present models where the cutoff is provided by the compositeness of $S$ or new fermions and collective symmetry breaking.

\subsection{One-loop corrections}

The scalar potential receives quantum corrections. At the one-loop level, the Coleman-Weinberg (CW) potential~\cite{Coleman:1973jx} under the $\overline{\rm MS}$ renormalization scheme is
\begin{align}
    \label{eq:CW potential}
    V_{\rm CW} = \frac{1}{64\pi^2} \left(\sum_B n_B\left( \log \left(\frac{m_B^2(h,S)}{Q^2}\right) - c_B\right) - \sum_F n_F \left(\log \left(\frac{m_F^2(h,S)}{Q^2} \right)- c_F\right) \right),
\end{align}
where $B,F$ is for bosons and fermions, and $n_{B,F}$ is the degrees of freedom. The constant $c_B = 3/2$ for longitudinal gauge bosons and scalar bosons, and $c_B = 1/2$ for transverse gauge bosons.
For fermions, $c_F = 3/2$. $Q$ is the renormalization scale.

The gauge and top Yukawa couplings are renormalized under the $\overline{\rm MS}$ renormalization scheme at $Q=m_Z = 91.1876~\gev$. The input parameters to compute those parameters are $m_W = 80.379~\gev$, $m_t = 172.69~\gev$, $\alpha_3 = 0.1184$, $m_h = 125.13~\gev$, and $G_F = 1.1663787\times 10^{-5}~\gev^{-1/2}$. The one-loop $\overline{\rm MS}$ parameters are computed following the expressions in Ref.~\cite{Buttazzo:2013uya}.
The quantum corrections from the couplings involving $S$ are negligible because of the weak coupling.
The parameters at a benchmark point $f = 5~\tev$, $\delta = 2\pi/5$, $m_S = 5~\gev$, and $\sin \theta = 0.1$ is
\begin{align}
    \label{eq:1-loop benchmark}
    A = 6.69~\gev,~\mu_S^2 = 181.51~\gev^2,~\mu_H^2 = 8765.52~\gev^2,~\lambda = 0.1495.
\end{align}

\begin{figure}[!t]
    \centering
    \includegraphics[width=0.6\linewidth]{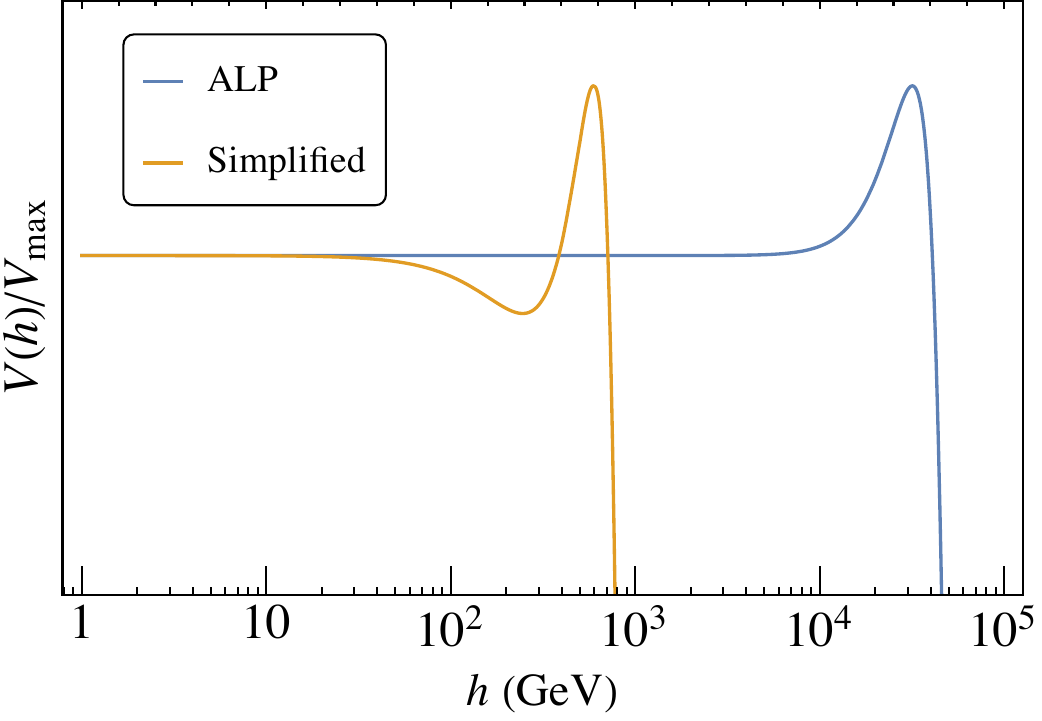}
    \caption{Comparison of the potential $V(h)$ along the path $\partial V/\partial S = 0$ between the full ALP model and the simplified model. The potential value is normalized by the value at the peak $V_{\rm max}$ of each one, about $10^{15}~\gev^4$ for the ALP model while $10^{7}~\gev^4$ for the simplified model.}
    \label{fig:T0 compare}
\end{figure}

The large negative contribution from the top quark in Eq.~\eqref{eq:CW potential} may destabilize the EW vacuum.
The interaction between $h$ and $S$ makes the potential of $h$ along the valley flatter than the SM. The quantum correction from the top then makes the potential energy turn down at a field value of $h$ not much above the EW scale. We, however, found that the EW vacuum is stable enough.
A comparison can be made with the simplified model, whose stability was confirmed in Ref.~\cite{Harigaya:2022ptp}.
A small effective quartic coupling along the ``valley'' appears in the simplified model so that the effective potential at $T=0$ starts to turn down at around $h=500~\gev$, potentially leading to the instability problem of the EW vacuum.
However, the tunneling rate from the EW vacuum to large field values through the barrier is suppressed by the large tunneling action $S_4$ due to the scalar kinetic energy;
from Eq.~\eqref{eq:Spath}, one can see that the field value shift of $S$ is huge for light ALPs, which contributes extra large kinetic energy to the tunneling action. Such suppression also happens in thermal phase transition, as discussed in Sec.~\ref{sec:nucleation}.
This large tunneling action protects the stability of the EW vacuum.
In the full ALP model, the path $\partial V/\partial S = 0$ is less flat because of the cosine coupling between $S$ and $H$ rather than a linear one, pushing the turning point where the effective potential starts to drop to a much higher field value, roughly 2 order of magnitude higher than that in the simplified model.
For example, in Fig.~\ref{fig:T0 compare}, we show the comparison of the potential as a function of $h$ along $\partial V/\partial S=0$ path between the ALP model and the simplified model at the benchmark point in Eq.~\eqref{eq:1-loop benchmark}.
The barrier between the metastable EW vacuum and the infinity appears around $h\simeq 600~\gev$ in the simplified model, while  $h\simeq 30~\tev$ in the full model.
The peak height is about $10^{15}~\gev^4$, compared to $10^{7}~\gev^4$ of the simplified model.
The $S$ field value in the ALP model is similar to the simplified model at least until the escape point of the simplified model.
In conclusion, like the simplified model, the tunneling action of the ALP model is large because of the large change of the $S$ field value,
and the high potential barrier and large change of the Higgs field value will further increase the tunneling action, resulting in an EW vacuum even more stable than the simplified model.

\section{The electroweak phase transition}
\label{sec:phase transition}

\subsection{Finite temperature corrections}

In addition to Eq.~\eqref{eq:CW potential},
the scalar potential receives finite-temperature correction term when $T > 0$
(see~\cite{Quiros:1999jp} for a review)\footnote{Notice that the sign convention in $J_F$ and $n_F$ is different from~\cite{Quiros:1999jp}.}
\begin{align}
    \label{eq:FT}
    V_{\rm FT}   & = \frac{T^4}{2\pi^2}\left(\sum_B n_B J_B\left(\frac{m_B^2(h,S)}{T^2}\right) + \sum_F n_F J_F\left(\frac{m_F^2(h,S)}{T^2} \right)\right),\nonumber \\
    J_{B,F}(x^2) & = \pm \int_0^\infty dy\ y^2 \log (1 \mp \exp (-\sqrt{y^2 + x^2})).
\end{align}

When the mass is not large compared with the temperature, we can use the high-temperature expansion of the integral function $J_{B,F}$.
This approximation is not well-justified for a strong first-order phase transition, but it provides qualitative information.
Here, we apply the high-temperature expansion as a quick, qualitative look for the finite temperature effective potential, while for the numerical computations we always use the full expressions.
Under the high-temperature expansion, the total effective potential becomes
\begin{align}
    \label{eq:high-T potential}
    V_{\rm highT} & = -\frac{1}{2}(\mu_H^2 - A f \cos \delta)h^2 + D_{\rm SM} T^2 h^2 - E_{\rm SM} T h^3 - E(h,S) T + \frac{1}{4}\lambda h^4 \nonumber                                 \\
                  & - \frac{1}{2} A f \cos \left(\frac{S}{f} - \delta\right) \left(\left(h^2 - 2v^2 + \frac{1}{3}T^2\right) - \frac{T^2}{24 f^2}\left(h^2-2v^2\right)\right) \nonumber \\
                  & - \left(1 - \frac{T^2}{24f^2}\right)\mu_S^2 f^2\cos\left(\frac{S}{f}\right),
\end{align}
where the Standard Model coefficients are
\begin{align}
    D_{\rm SM} & = \frac{1}{32}\left(3 g^2 + g_1^2 + 4 y_t^2 + 8\lambda\right), \nonumber \\
    E_{\rm SM} & = \frac{1}{32\pi}\left(2g^3 + (g^2+g_1^2)^{3/2}\right).
\end{align}
Among the corrections from the couplings involving $S$,
the term proportional to $T^2/3$ in the second line of Eq.~\eqref{eq:high-T potential} is the dominant effect.
The term $E(h,S)$ comes from the IR singularity of the scalar boson contribution, as $E_{\rm SM}$.
The exact form is complicated but roughly behaves as $(A f \cos (S/f - \delta))^{3/2}$.
This term contributes positively to the phase transition strength, but is smaller than or at most comparable with other terms and does not have a qualitative impact on the potential shape.
We thus neglect this term in our high-temperature, analytical discussion. (Full scalar contributions are included in the numerical computations).
The term proportional to $T^2/f^2$ in the second and third lines is also small due to the large $f$ and thus can be neglected. Under these approximations, the high-temperature expanded potential becomes
\begin{align}
    \label{eq:highT simplified}
    V_{\rm highT} & = -\frac{1}{2}(\mu_H^2 - A f \cos \delta)h^2 + D_{\rm SM} T^2 h^2 - E_{\rm SM} T h^3 + \frac{1}{4}\lambda h^4 \nonumber \\
                  & - \frac{1}{2} A f \cos \left(\frac{S}{f} - \delta\right) \left(h^2 - 2v^2 + \frac{1}{3}T^2\right) \nonumber             \\
                  & - f^2\mu_S^2 \cos\left(\frac{S}{f}\right).
\end{align}
Even for the case where the gauge bosons and fermions are not light compared with the temperature, Eq.~\eqref{eq:highT simplified} can provide a qualitative intuition.
The thermally generated $h^3$ barrier is the same as the SM (up to the additional subdominant $E(h,S)$ term), but the enhancement of the EWPT strength can come the zero-temperature potential, as discussed in the next subsection.

The ``valley'' now becomes
\begin{align}
    \label{eq:valley FT}
    S(h,T) = f \arctan \left(\frac{A (3(h^2 - 2v^2) + T^2) \sin(\delta)}{6f \mu_S^2 + A(3(h^2 -2v^2) + T^2) \cos(\delta)}\right).
\end{align}
Again, one can retrieve the path in the simplified model by taking $f\rightarrow \infty$,
\begin{align}
    \label{eq:valley large f finite T}
    S (h, T) = \frac{1}{2} \frac{A}{\mu_S^2}(h^2 - 2v^2 + \frac{1}{3}T^2) \sin \delta.
\end{align}
This shows that for any high temperature $T$ and Higgs field $h$, $S$ always has a single minimum. When the Higgs field vev is fixed at $\langle h \rangle = 0$ at high temperature, the global vev goes to $(0, S (0, T))$. Thus the phase transition is a simple one-step one with $(0,S (0, T))\rightarrow (v(T), S (v(T), T))$.

In addition to \cref{eq:full model physical,eq:CW potential,eq:FT}, we need to include the ring diagram contribution to resum the effective potential in order to keep the validity of the perturbative computation. This requires adding the thermal corrections to the boson masses in \cref{eq:CW potential,eq:FT}. The one coming from the gauge bosons is the same as the SM result, i.e.,
\begin{align}
    \label{eq:gauge thermal mass}
    m_{A_L,Z_L}^2 = & \begin{pmatrix}
                          \frac{1}{4}g^2 h^2 + \frac{11}{6}g^2 T^2 & - \frac{1}{4} g g_1 h^2                      \\
                          - \frac{1}{4} g g_1 h^2                  & \frac{1}{4}g_1^2 h^2 + \frac{11}{6}g_1^2 T^2
                      \end{pmatrix}, \nonumber \\
    m_{Z_T}^2 =     & \frac{1}{4}(g^2+g_1^2)h^2, \nonumber                                                    \\
    m_{A_T}^2 =     & 0, \nonumber                                                                            \\
    m_{W_L^\pm} =   & \frac{1}{4}g^2 h^2 + \frac{11}{6}g^2 T^2, \nonumber                                     \\
    m_{W_T}^2 =     & \frac{1}{4} g^2 h^2,
\end{align}
where the subscript $L$, $T$ are for longitudinal and transverse modes, respectively.
The $A_L$ and $Z_L$ boson masses are the eigenvalues of the mass matrix $m_{A,Z}$. Note that only the longitudinal mode of the gauge bosons receives the thermal mass corrections $\Pi \sim g^2 T^2$. The thermal mass matrix for the scalar bosons can be derived from Eq.~\eqref{eq:highT simplified}.
The $(1,1)$ and $(2,2)$ component of the mass matrix in Eq.~\eqref{eq:scalar mass matrix} receives thermal mass corrections,
\begin{align}
    \label{eq:scalar thermal mass}
    \Pi_{hh} = 2 D_{\rm SM} T^2, \Pi_{SS} = \frac{A}{2f} \cos (S/f-\delta) \times \frac{1}{3} T^2.
\end{align}
The thermal correction $\Pi_{hh}$ also applies to the would-be Nambu-Goldstone bosons.

Throughout this paper, all numerical computations for the thermal phase transition are performed following the above computed one-loop thermal effective potential with resummation, i.e., \cref{eq:full model physical,eq:CW potential,eq:FT} with \cref{eq:gauge thermal mass,eq:scalar thermal mass} included.
Again, high-temperature expansion is never employed in the numerical computations.

\subsection{Phase transition and bubble nucleation}
\label{sec:nucleation}

\begin{figure}[!t]
    \centering
    \includegraphics[width=0.5\linewidth]{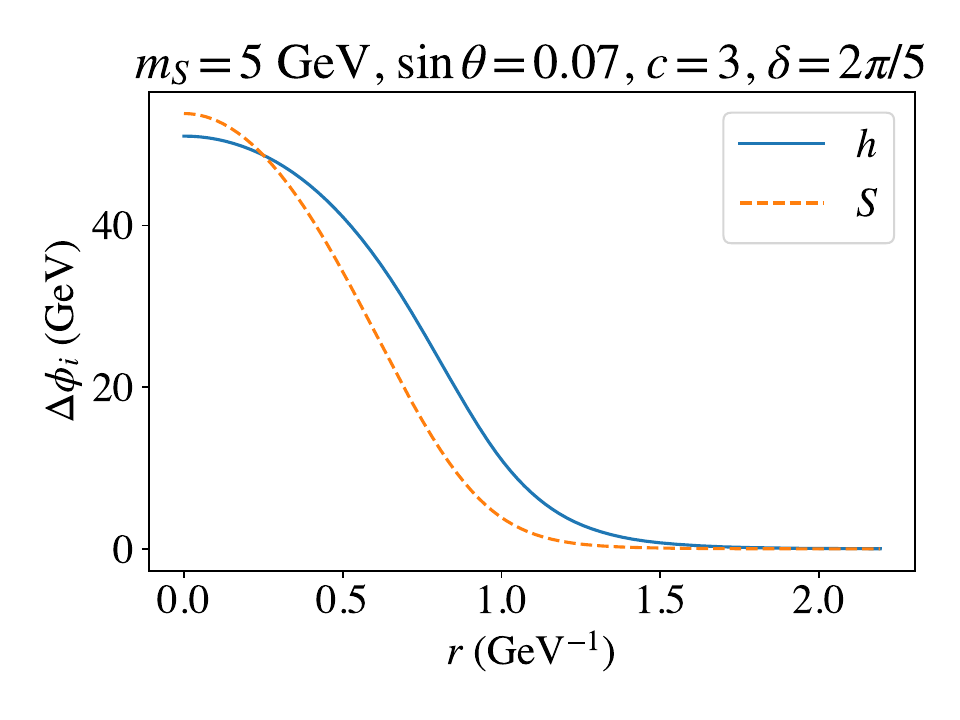}
    \caption{Wall profile during phase transition for a benchmark point $m_S = 5~\gev$, $\sin \theta = 0.07$, $c=3$ and $\delta=2\pi/5$. The change of the field value $\Delta \phi_i = \Delta h, \Delta S$, is defined so that they are $0$ for $r \rightarrow \infty$.}
    \label{fig:profile}
\end{figure}

At the critical temperature $T_c$, the thermal potential is degenerate for the false vacuum $(0, S (0,T_c))$ and the true vacuum $(v(T_c), S (v(T_c),T_c))$.
As the universe continues cooling down, a thermal phase transition from the false vacuum to the true vacuum is kinetically allowed.
Once the transition rate is larger than the Hubble rate, first-order phase transition proceeds via bubble nucleation. The criteria for such a condition is~\cite{Quiros:1999jp,Linde:1981zj}
\begin{align}
    \label{eq:nucleation condition}
    \left. \frac{S_3}{T}\right\vert_{T_n}  \simeq 140.
\end{align}
The temperature $T_n$ when such a condition is satisfied is referred to as the nucleation temperature. The 3-dim Euclidean action, $S_3$, is the sum of the potential energy and the kinetic energy of both scalar fields,
\begin{align}
    \label{eq:Euclidean action}
    S_3 = 4\pi \int r^2 \left(\frac{1}{2}\left( \frac{d h(r)}{dr} \right)^2 + \frac{1}{2} \left( \frac{d S(r)}{dr} \right)^2  + V(h(r),S(r))\right).
\end{align}
Here $r$ is the space coordinate and the integral is performed along the bubble profile whose $h(r)$ and $S(r)$ minimize such an action.
Numerical computation shows that the path (i.e., the profile) of the thermal phase transition is always almost along the ``valley'' where $\partial V/\partial S = 0$. In Fig.~\ref{fig:profile}, we show the field value shift of $h$ and $S$ for a benchmark point $m_S = 5~\gev$, $\sin \theta = 0.07$, $c=3$, and $\delta=2\pi/5$.

\begin{figure}[!t]
    \centering
    \includegraphics[width=0.465\linewidth]{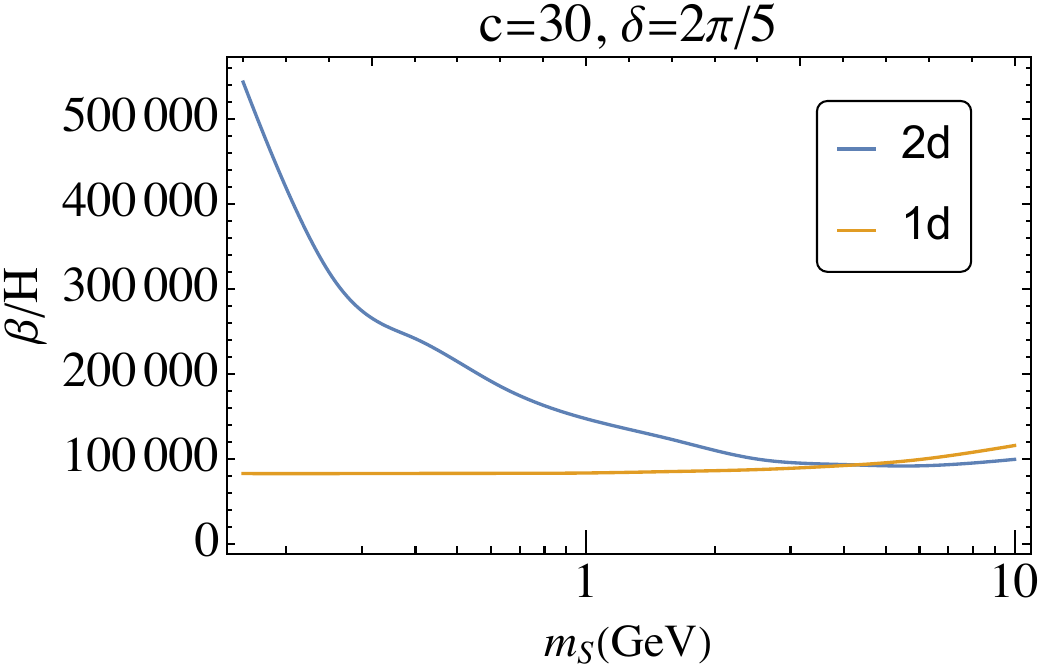}
    \includegraphics[width=0.435\linewidth]{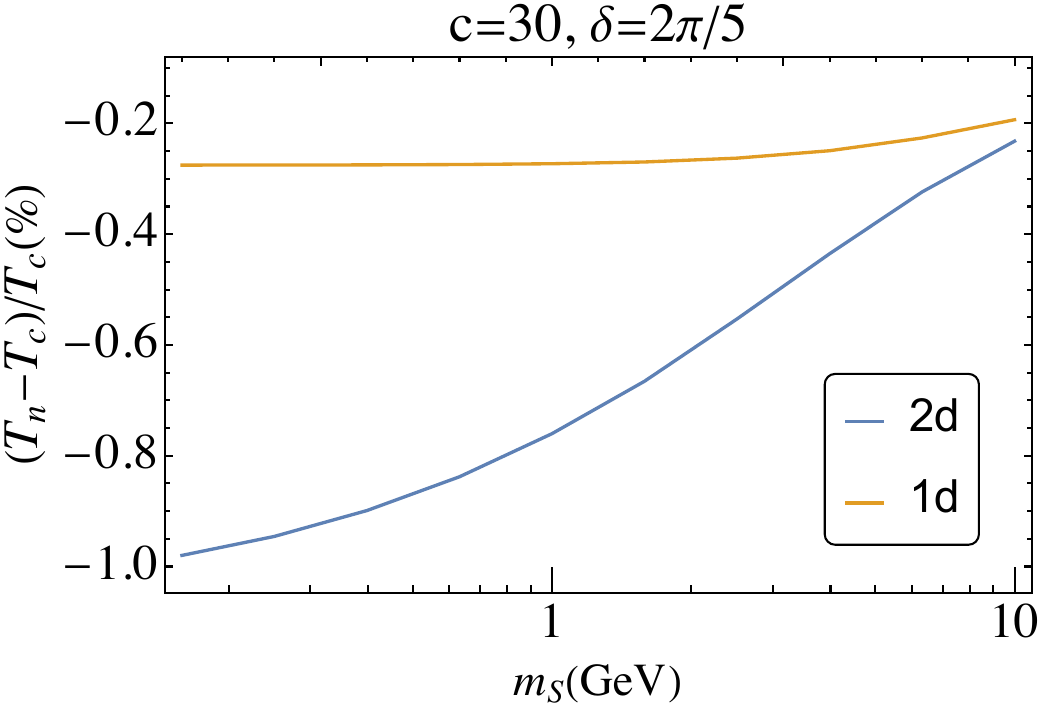}
    \caption{
        Properties of the phase transition. The mixing angle $\theta$ is fixed so that $m_S^2/\mu_S^2=0.13$. ``2d" is the result of the full two-field dynamics.
        ``1d" is the result of the approximation where the path is fixed on the valley of potential and  the kinetic term of $S$ is neglected.
        Left: The inverse duration of the phase transition, $\beta/H$. One can see that it blows up for small $m_S$.
        Right: the difference between nucleation temperature $T_n$ and the critical temperature $T_c$ normalized by $T_c$. One can see that the nucleation temperature is more delayed for smaller $m_S$, but the delay is less than $1\%$ and the increase of the delay gradually slows down for the MeV scale $m_S$.}
    \label{fig:1d and 2d nucleation}
\end{figure}

From Eq.~\eqref{eq:valley large f finite T}, one can see that the field value shift of $S$  is larger for a smaller scalar mass. Such a large field value shift will contribute huge kinetic energy to Eq.~\eqref{eq:Euclidean action}, as discussed in the previous paper~\cite{Harigaya:2022ptp}.
Another important quantity is the inverse time duration, defined as
\begin{align}
    \label{eq:betaH}
    \frac{\beta}{H} \equiv T_n \left.\frac{d}{dT} \left( \frac{S_3}{T} \right) \right \vert_{T_n}.
\end{align}
In the left panel of Fig.~\ref{fig:1d and 2d nucleation}, we show $\beta/H$ as a function of $m_S$. The mixing angle is derived by fixing the tuning parameter $m_S^2/\mu_S^2 = 13\%$.
Numerical computations of the effective potential and thermal phase transition action are performed with the help of \verb|cosmoTransitions|~\cite{Wainwright:2011kj}.
Here ``2d" is the result of the full two-field dynamics, while ``1d" is the result of a truncated computation where the path is confined to the 1-dimensional valley and the kinetic term of $S$ is neglected.
One can see that the $\beta/H$ parameter increases quickly as $m_S$ decreases for the full 2-dim computation.
The truncated 1-dim computation significantly underestimates $\beta/H$ if $m_S <10$ GeV.

$\beta/H$ is large for the following reason. Because of the weak coupling of $S$, each term in $S_3$ is large.
Around the nucleation temperature, the large positive contribution from the kinetic energy of $S$ and $h$ cancels with the large negative contribution from the potential, and thus $S_3/T \simeq 140$ is achieved. However, this cancellation does not occur for their derivatives, causing huge $\beta/H$.
The large $\beta/H$ parameter has significant impacts.
A large $\beta/H$ corresponds to a short time duration of the phase transition, which suppresses the gravitational-wave signals.

A large $\beta/H$ also implies a rapid decrease of the tunneling action $S_3$ as the temperature drops. Such a rapid change compensates for the large $S_3$ coming from the huge field value shift of $S$, preventing the nucleation temperature from being much below the critical temperature.
In the right panel of Fig.~\ref{fig:1d and 2d nucleation}, we show the comparison between $T_n$ and $T_c$ as a function of $m_S$.
$T_n$ deviates more from $T_c$ for the 2-dim computation, but the difference is still below one percent.

A SFOPT is defined as a FOPT such that the sphaleron processes decouple in the broken phase, i.e., the sphaleron rate at the $T_n$ in the broken phase is smaller than the Hubble rate.
For the computation of the sphaleron rate, see~\cite{Moore:1998swa,DOnofrio:2014rug}. Rather than the commonly used SFOPT criteria $v_c/T_c \geq 1$, the physical criteria for SFOPT should be $v_n/T_n \geq 1$, and sometimes they can favor very different parameter spaces~\cite{Baum:2020vfl}. In this model, however, we numerically found that
\begin{align}
    \label{eq:strength Tn Tc}
    \frac{v_n}{T_n} \simeq 1.2 \frac{v_c}{T_c}.
\end{align}
To speed up the computation and avoid numerical fluctuations, we numerically compute $v_c/T_c$, use Eq.~\eqref{eq:strength Tn Tc} to estimate $v_n/T_n$, and then require $v_n/T_n \geq 1$ for the SFOPT. Prediction on the minimal $\sin \theta$ for a certain $m_S$ for SFOPT under $v_c/T_c \geq 1$ and $v_n/T_n \geq 1$ differ from each other by less than $10\%$.

\begin{figure}[!t]
    \centering
    \includegraphics[width=0.325\linewidth]{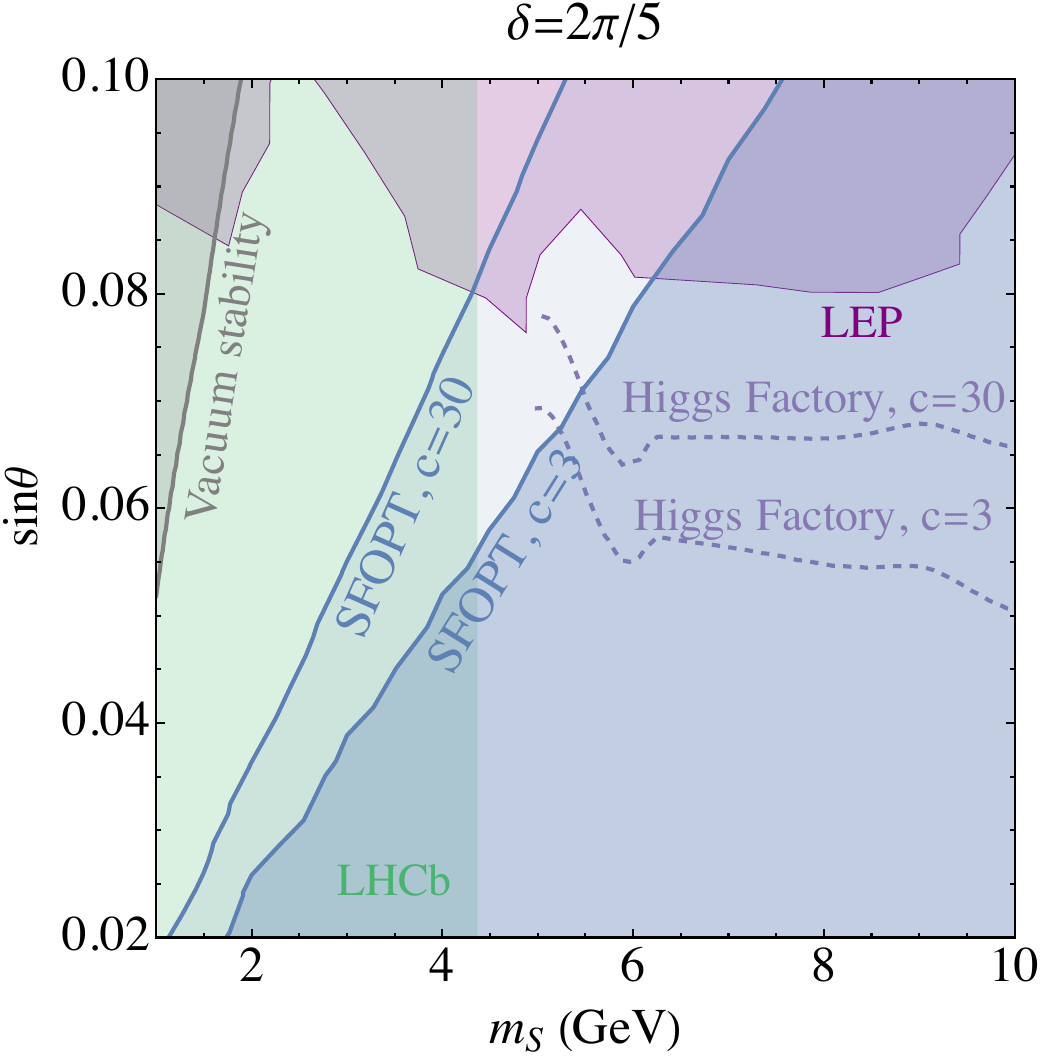}
    \includegraphics[width=0.325\linewidth]{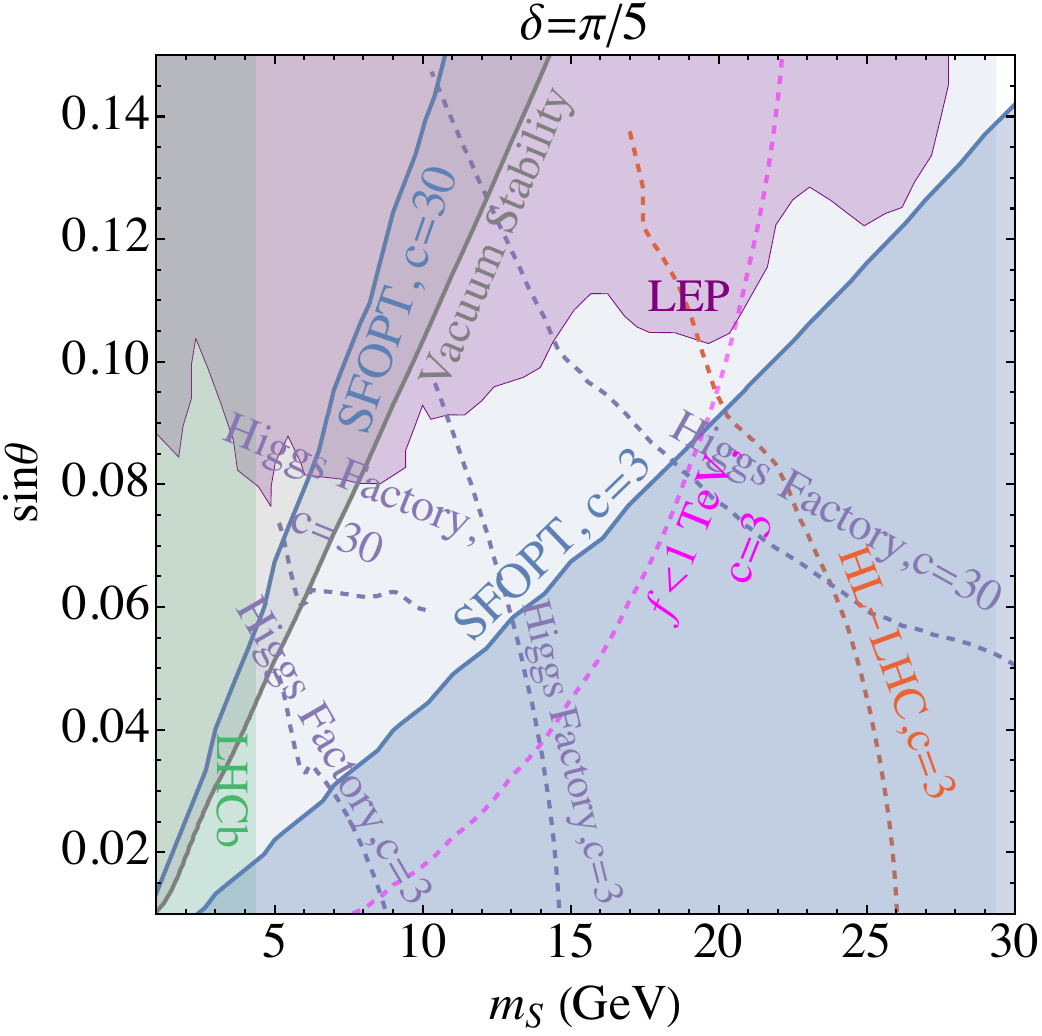}
    \includegraphics[width=0.325\linewidth]{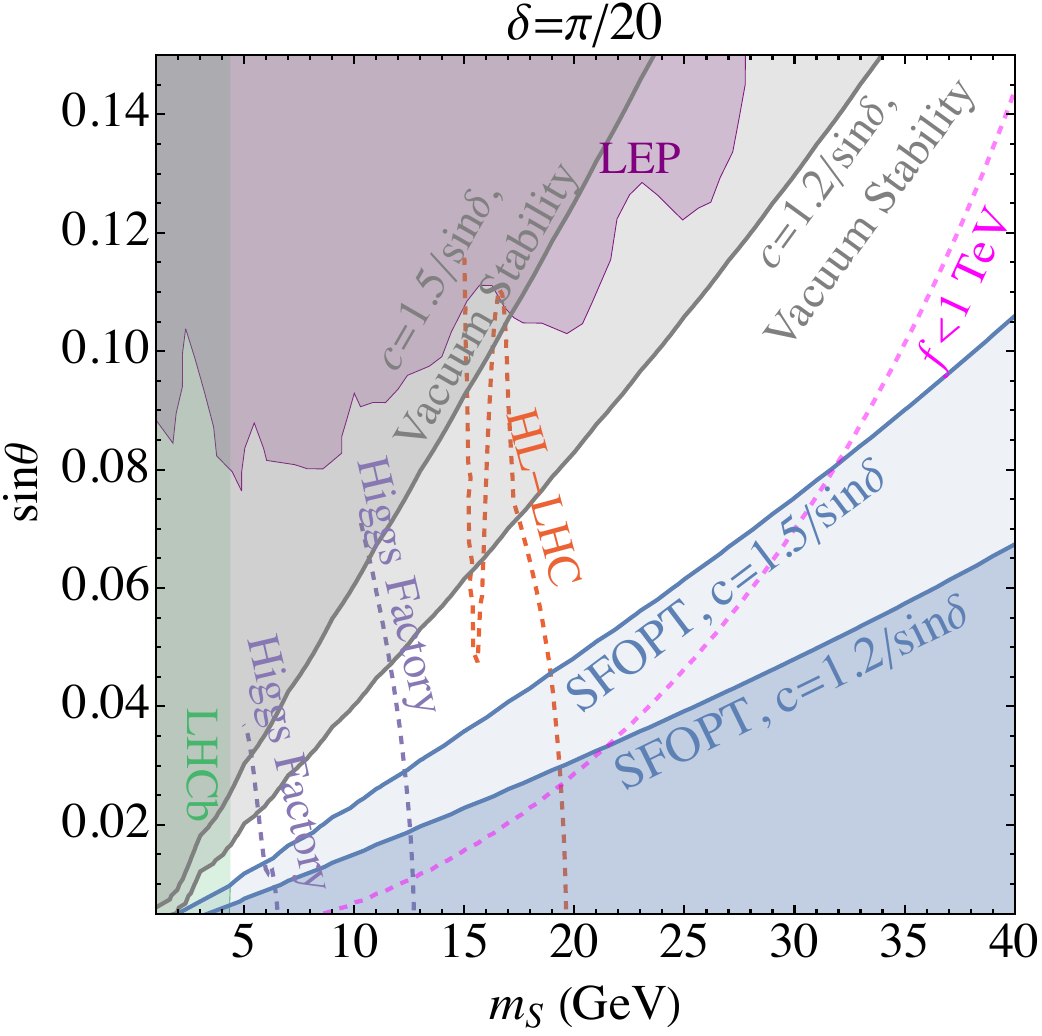}

    \vspace{0.3cm}

    \includegraphics[width=0.325\linewidth]{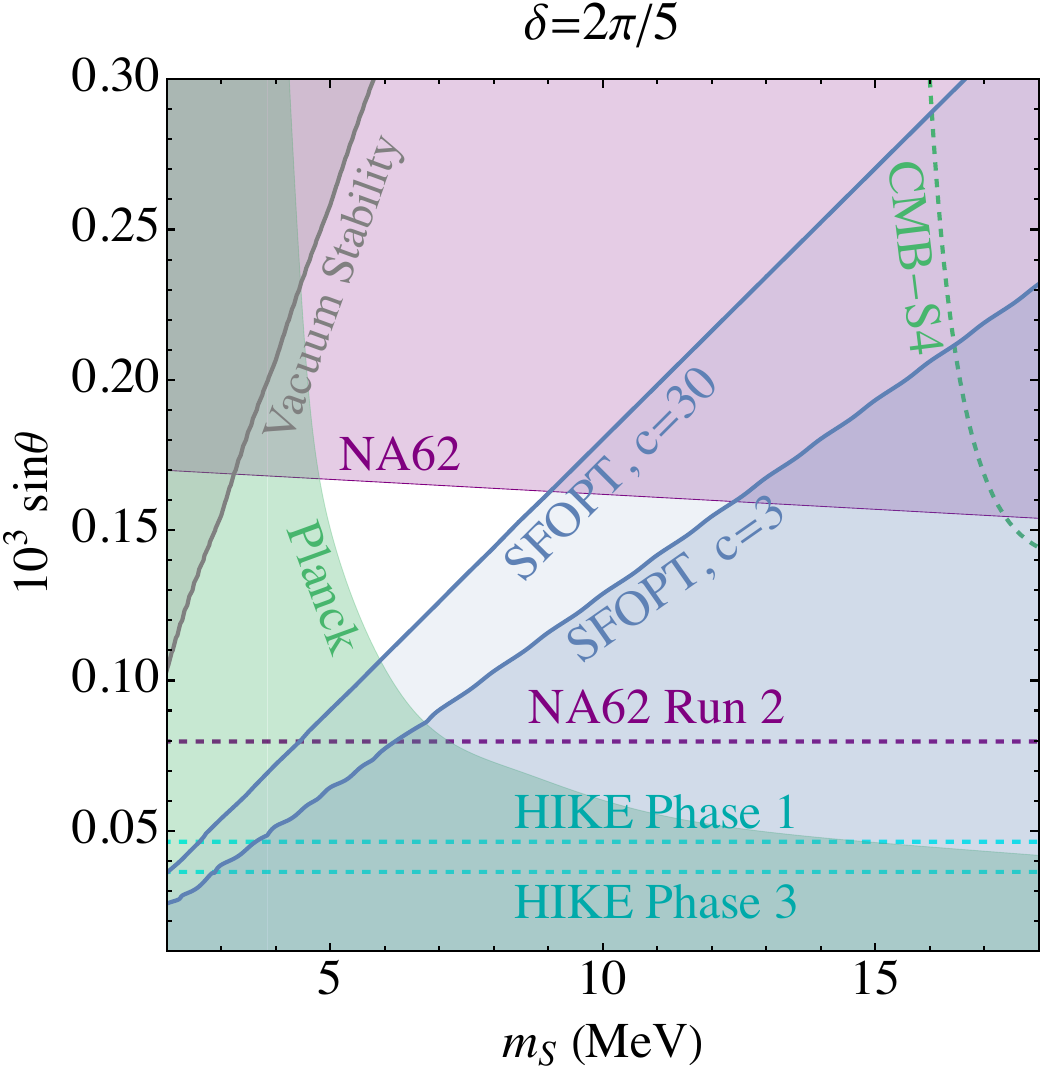}
    \includegraphics[width=0.33\linewidth]{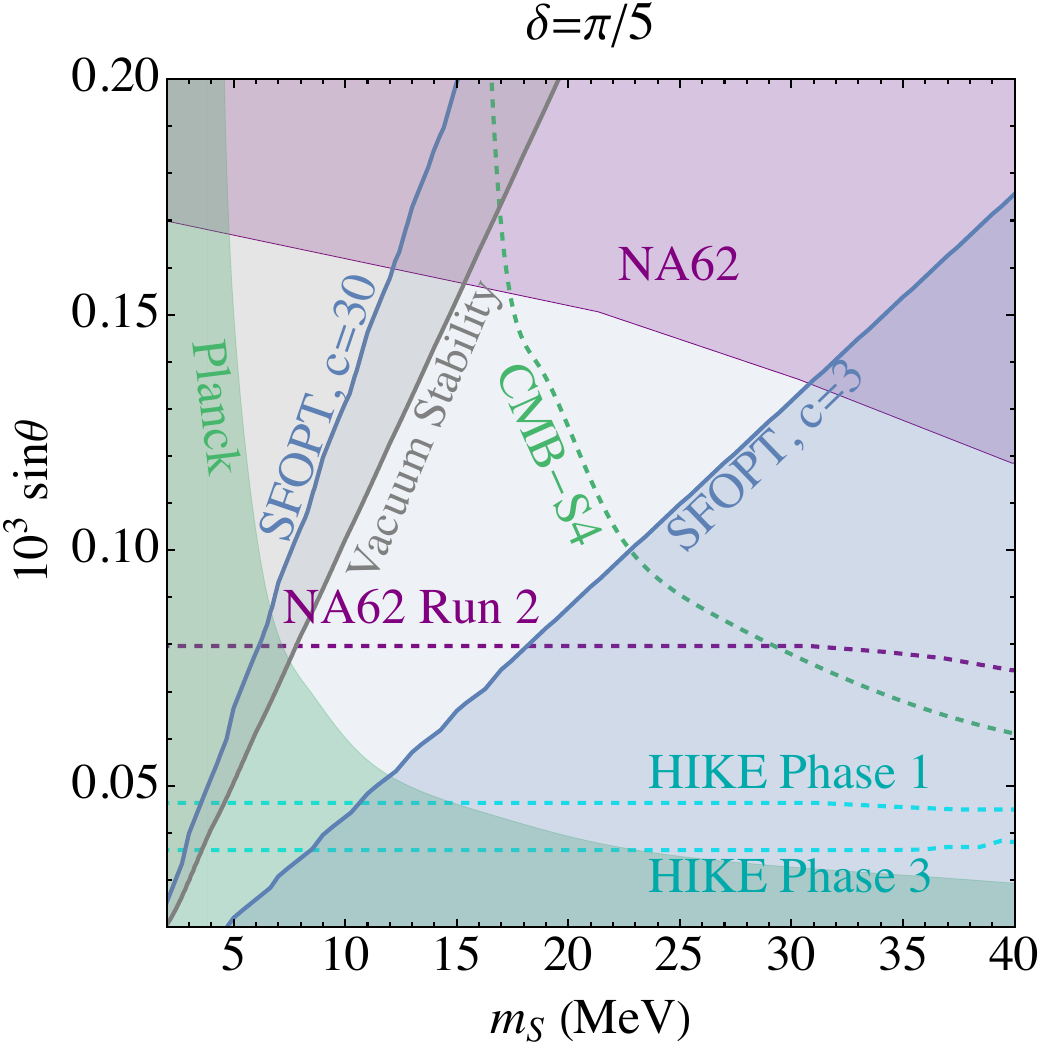}
    \includegraphics[width=0.325\linewidth]{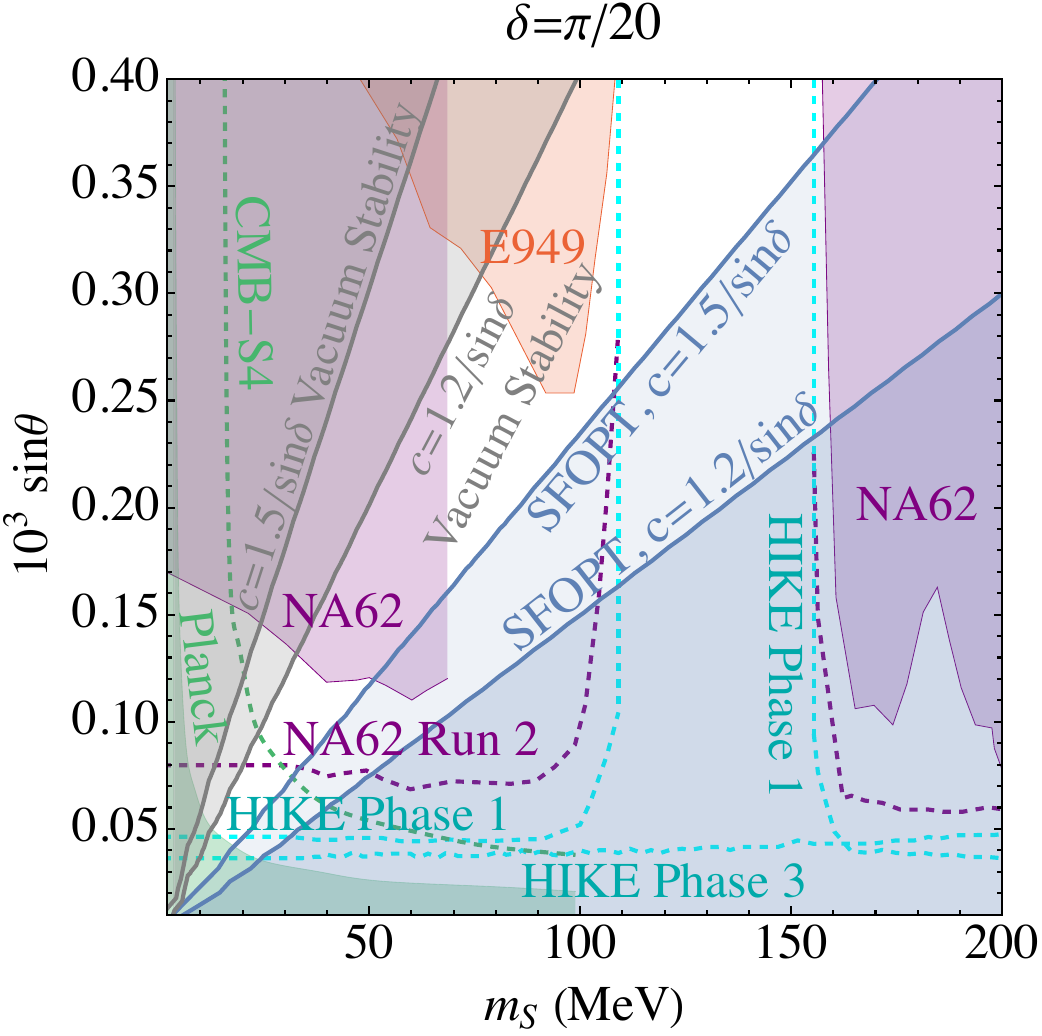}
    \caption{Parameter space in $m_S - \sin \theta$ plane for different $c$ and $\delta$. Upper panels: $m_S$ in the GeV scale. For $\delta=\pi/20$, the Higgs exotic search projection curve and $f=1~\tev$ curve for $c=1.2/\sin \delta$ and $c=1.5/\sin \delta$ are very close to each other and we only show the curves for $c=1.2/\sin \delta$. Lower panels: $m_S$ in the MeV scale. Gray shaded region: excluded by vacuum stability requirement $V(0) > V(\sqrt{2}v)$. For $\delta=\pi/5$ or $2\pi/5$, this requirement is only plotted for $c=3$. The curve for $c=30$ is outside the plot range.}
    \label{fig:final result}
\end{figure}

In Fig.~\ref{fig:final result}, we showed the bound on the model for
several choices of $(c,\delta)$.
In the blue-shaded regions,
the SFOPT is not achieved.
Computation and numerical scanning for the critical temperature are performed with a Python code, and cross-checked with a \verb|Mathematica| code for a few parameter points.
Among them, $c=3, \delta=\pi/5$ and $c=3, \delta=2\pi/5$ requires
$\Lambda_H \lesssim$ few $\tev$.
Increasing $c$ by $10$ times for fixed $\delta$ relaxes this upper bound for $\Lambda_H$ by a factor of $\sqrt{10}$.
Via numerical computation, we find that the ratio $\Delta S/f$ during phase transition becomes smaller for $c<3$, which in turn leads to a weaker PT strength, thus we do not explore smaller $c$ for $\delta=2\pi/5, \pi/5$.
For smaller $\delta$, $c=3$ violates the EW stability requirement in Eq.~\eqref{eq:ew-stability}.
For this reason, for $\delta = \pi/20$ we take $c = 1.2/\sin \delta$ and $c = 1.5/\sin \delta$ in Fig.~\ref{fig:final result}.
These values of $c$ require $\Lambda_H \lesssim $ few $\tev$.
As $\delta$ becomes smaller with $(m_S, \sin \theta, c \sin \delta)$ fixed, $v_c/T_c$ slowly increases up to by 10\%, but this does not change the lower bound on the mixing angle beyond the uncertainty of our computation discussed below.

Our computation includes the one-loop zero-temperature CW potential and the one-loop finite temperature corrections with resummation. Including the CW correction is important. Indeed, if we remove the CW potential and use the tree-level zero-temperature potential, $v_c/T_c$ increases by an $O(1)$ factor and reach even $\sim 3$. The prediction on the mixing angle changes correspondingly.

However, this computation method, though commonly used in this community, still suffers from large computational uncertainties. As discussed in the Appendix, higher-loop thermal corrections increase the prediction on $v_c/T_c$ in the SM with a lighter Higgs by as much as 50\% in comparison with the one-loop computation. A similar enhancement of the PT strength may happen in the ALP model. A careful two-loop computation is required in the future to make a more convincing prediction but is beyond the scope of the current paper. (The publicly available code \verb|DRalgo| package~\cite{Ekstedt:2022bff} cannot treat cosine potentials.)
If $v_c/T_c$ is actually larger than the prediction of the one-loop computation by 50\%, the predicted SFOPT lower bound on $\sin\theta$ for each ALP mass may be relaxed by about 20\% for $c\sin \delta = O(1)$.

\subsection{Enhancement of the phase transition}
\label{sec:PT enhance}

To see how EWPT strength is enhanced in comparison with the SM and the simplified model,
we work up to order $A f^{-1} S^2 h^2 \cos \delta$. We see that the ``valley'' at $T=0$ is now expressed as
\begin{align}
    \label{eq:Spath h-2}
    S & = \frac{A f (h^2 - 2v^2) \sin \delta}{2 f \mu_S^2 + A (h^2 - 2 v^2) \cos \delta} \nonumber                                                                                                                                                    \\
      & \simeq \frac{A f v^2 \sin \delta}{A v^2 \cos \delta - f \mu_S^2} + \frac{A f^2 \mu_S^2 \sin \delta}{2 (A v^2 \cos \delta - f \mu_S^2)^2}h^2 + \frac{A^2 f^2 \mu_S^2 \sin\delta \cos \delta}{4 (A v^2 \cos \delta - f \mu_S^2)^3}h^4 + O(h^6).
\end{align}
Along this path, the 1-dim tree-level potential has a positive $h^6$ term, while the quartic term is
\begin{align}
    \label{eq:tree-quartic-1d}
    \frac{1}{4}(\lambda - \frac{A^2 f^3 \mu_S^4 \sin^2 \delta}{2(\mu_S^2 f - A v^2 \cos \delta)^3}) h^4.
\end{align}
Since we assume $\mu_S^2 f > A v^2$, the correction always makes this effective quartic coupling smaller than $\lambda$.
However, when large $A$ or small $f$ is taken so that $\mu_S^2 f$ is not much above $A v^2$, the denominator of the correction term can be small so that the large correction may even make the effective quartic coupling negative.
This negative $h^4$ term contributes an extra barrier in addition to the traditional thermal $h^3$ barrier, while the local minimum comes from the competition between the negative $h^4$ and positive $h^6$. This significantly enlarges the barrier and thus enhances the strength of EWPT. With numerical computations, we found that this is indeed the case in the parameter region where SFOPT is achieved.

The model requires mild tuning.
After we express $f$ in terms of $c f_c$ following Eq.~\eqref{eq:fc def}, the correction in Eq.~\eqref{eq:tree-quartic-1d} still behaves as $A^2/\mu_S^2 \propto (m_h^2 - \mu_S^2)(1-m_S^2/\mu_S^2)$. A smaller $m_S^2/\mu_S^2$ helps make the effective quartic coupling negatively large and achieve SFOPT. Small $m_S^2/\mu_S^2$ requires mild tuning of parameters. For example, the benchmark point in Eq.~\eqref{eq:tree-level benchmark} has $m_S^2/\mu_S^2=0.14$.
The SFOPT boundary in the parameter space is almost parallel with a contour for a fixed $m_S^2/\mu_S^2$, depending on the $c$ and $\delta$. For example, the SFOPT boundary for $c=3$ and $\delta = 2\pi/5$ has roughly $25\%$ tuning.

\section{Local electroweak baryogenesis}
\label{sec:ewbg}

In local electroweak baryogenesis, the CP violation and B violation occur at the same local spacetime during the electroweak phase transition as the wall passes through. We first briefly review the mechanism in a general way, following Ref.~\cite{Dine:1990fj,Dine:1992ad}.
Consider an operator $\hat{\mathcal{O}}  = \alpha_2 \theta W \tilde{W}/(8 \pi)$ or $\hat{{\mathcal{O}}} = \partial_\mu \theta \psi^\dagger \bar{\sigma}^\mu \psi$, where $W$ is the field strength of the $SU(2)$ gauge field, $\psi$ is an $SU(2)$ doublet fermion, and $\theta$ depends on some fields whose field values shift during the phase transition.
In the thick-wall regime where the wall thickness is much larger than the mean free path of the particles, $\theta$ changes slowly and we may assume a small departure from local thermal equilibrium.
The time-dependence of $\theta$ gives the imbalance between the sphaleron and anti-sphaleron rates via this operator, and the net baryon number is produced.
This can be understood by the effective chemical potential of the Chern-Simons number of $W$ or the fermion number of $\psi$, with an equilibrium value of the baryon number $n_{B}^0 \sim \dot{\theta}T^2$.
See~\cite{Domcke:2020kcp,Co:2020xlh} for the Boltzmann equation of the SM particles under $\dot{\theta}\neq 0$.
We may then determine the evolution of the baryon asymmetry from the detailed balance principle,
\begin{align}
    \dot{n}_B \propto -\Gamma_{\rm sph} (n_B - n_B^0).
\end{align}
where $\Gamma_{\rm sph} \simeq 20 \alpha_W^5 T$ is the sphaleron rate per volume.
Since the sphaleron rate is highly suppressed by $\alpha_W^5$, the produced baryon number $n_B$ is always much smaller than the $\dot{\theta} T^2$ and thus we can ignore $n_B$ on the right-hand side of this equation. Integrating over time, we obtain
\begin{align}
    n_B \propto \Gamma_{\rm sph} T^2\Delta \theta,
\end{align}
where $\Delta \theta$ is the shift of $\theta$ from outside the bubble wall to inside until the sphaleron rate is suppressed.

Local electroweak baryogenesis was considered for dim-6 operator $|H|^2 W \tilde{W}/M^2$ in Ref.~\cite{Dine:1990fj,Dine:1992ad}, where $M$ is a dimensionful parameter. Here, $\theta$ is identified with $|H|^2/M^2$.
However, as stated in the introduction, $M$ must be right above the weak scale because of the small field value shift of $h$.
On the other hand, in our model, we may instead consider the following dim-5 operators,
\begin{align}
    \label{eq:dim-5 operators}
    \mathcal{L} = \frac{\alpha_2}{8\pi} c_W \frac{S}{f} W \tilde{W},~\frac{\partial_\mu S}{f} c_{q_{ij}} q^\dagger_i \bar{\sigma}^\mu q_j,~\text{or}~\frac{\partial_\mu S}{f} c_{\ell_{ij}} \ell^\dagger_i \bar{\sigma}^\mu \ell_j,
\end{align}
where $q$ and $\ell$ are quark and lepton doublets, respectively. $c_W$, $c_{q_{ij}}$ and $c_{\ell_{ij}}$ are coupling constants.
We can take $c_{\psi_{ij}} = \delta_{ij} c_{\psi_i}$ by unitary rotations of fermion fields, where $\psi=q$ or $\ell$.
The coupling with $W$ is also discussed in Ref.~\cite{Jeong:2018ucz}.
These effective operators can be generated by UV physics as will be discussed in Sec.~\ref{sec:UV}.
The $SW\tilde{W}$ coupling is generated through the weak anomaly of the shift symmetry of $S$, while the $(\partial S)\psi^\dagger \bar{\sigma}\psi$ coupling is generated through a non-zero $U(1)$ charge of SM fermions or the mixing of SM fermions with $U(1)$-charged heavy fermions.
The large field value shift of $S$ during the phase transition means that the required value of $f$ for a successful baryogenesis is large, so that the electron EDM constraint is avoided.

Here, we follow the computation in Ref.~\cite{Domcke:2020kcp,Co:2020xlh} to derive the Boltzmann equation for $n_q$. Defining $n_{q_i}$ to contain all the colors of quark doublets for each generation, the part of the Boltzmann equation relevant for baryogenesis can be written as
\begin{align}
    \label{eq:bolz}
    \dot{n}_{q_i} = 9 \Gamma_{\rm sph} \sum_j \left(-n_{q_j} - n_{\ell_j} + \frac{3 c_{q_j} + c_{\ell_j} - c_W/3}{3} \dot{\theta} T^2\right),
\end{align}
where the overall coefficient $9$ is derived from the result in Ref.~\cite{Cline:1995dg}.
The baryon number is thus
\begin{align}
    \label{eq:baryon number}
    n_B = \Gamma_{\rm sph} T^2 \frac{\Delta S}{f}\sum_i(9 c_{q_i} + 3 c_{\ell_i} - c_W).
\end{align}

If $c_W = 1$ and $c_{q_i} = c_{\ell_i} = 0$, we find that the BAU produced in the viable parameter space is larger than the observed one by an $O(1 -10)$ factor for $c=1.5/\sin \delta$ with small $\delta$ and $c=3$ for $\delta=\pi/5$, and by an $O(20-50)$ factor for $c=1.2/\sin \delta$ with $\delta = \pi/20$. For larger $c$ and $\delta$, the produced BAU can be smaller and reach the observed amount. For example, the produced BAU is around the observed value for $c=3, \delta = 2\pi/5$.
For a model with $c_{q_i} = 1$ or $c_{\ell_i} = 1$ while $c_W = 0$, the produced baryon number is 9 or 3 times larger that the model with $c_W=1$ and $c_{q,\ell}=0$, respectively.
Explaining the observed BAU for those overproduced cases requires suppression of the operators in Eq.~\eqref{eq:dim-5 operators}.
The realization of the suppression in UV theories is discussed in the next section.

\section{UV completion}
\label{sec:UV}

In this section, we discuss the UV completion of the ALP-assisted model.
In Sec.~\ref{sec:composite}, we present models where the ALP $S$ arises by a spontaneous breaking in a QCD-like theory. The smallness of $f$ in comparison with other fundamental scales is explained by the dimensional transmutation. As we will see, the model can be UV-completion for $m_S= O(10)$ GeV.
In Sec.~\ref{sec:perturbative}, we discuss a model where $S$ is a phase direction of a fundamental complex scalar field. The model can be a UV completion for both $m_S = O(10)$ MeV and GeV, but does not explain the smallness of $f$ in comparison with other fundamental scales. To explain it, the model should be further extended, e.g., with supersymmetry with a supersymmetry breaking scale at or below $f$. For $m_S = O(10)$ MeV, where $f = O(10^{6-7})$ GeV, this does not require low-scale supersymmetry.
See Ref.~\cite{Jeong:2018ucz} for another UV-complete model.

\subsection{Composite UV completion}
\label{sec:composite}

We first discuss a model where $SW\tilde{W}$ coupling is $\alpha_2/(8\pi f)$-suppressed without further suppression.
The model is consistent with the parameter region where the observed BAU can be explained if $c_W \simeq 1$.
We introduce an $SU(N)$ gauge interaction ($N\geq 3$) with fermions shown in Table~\ref{tab:composite-fermion2}. The $SU(N)$ theory has 3 flavors, and the $SU(3)_L\times SU(3)_R$ global symmetry is broken down into $SU(3)_V$, yielding the following 8 NGBs:
\begin{align}
    T(3,0),~ D(2,1/2),~ S(1,0),
\end{align}
where the numbers in the parenthesis denote the $SU(2)_L\times U(1)_Y$ charges.
There is one neutral NGB $S$, and the corresponding spontaneously broken global $U(1)$ charges of the fermions are also shown in Table~\ref{tab:composite-fermion2}. $U(1)$ has $SU(2)_L\times U(1)_Y$ anomaly, so $S$ has  anomalous couplings to the electroweak gauge bosons suppressed by $\alpha_2/(8\pi f)$.
Non-zero masses of $L\bar{L}$ and $E\bar{E}$ give a non-zero mass to $S$.

We introduce the following couplings with the Higgs;
\begin{align}
    \label{eq:yukawa}
    {\cal L}= y H L \bar{E} + \bar{y} H^\dag \bar{L} E + {\rm h.c.}
\end{align}
These couplings violate $U(1)$ and generate $H-D$ mass mixing whose phase depends on $S$. The exchange of $D$ generates the couplings of $S$ with the Higgs,
\begin{align}
    \label{eq:SH_UVC}
    \frac{\Lambda_N^4}{16\pi^2}\frac{y \bar{y}}{ m_{D}^2} e^{i S/f} |H|^2  + {\rm h.c.},
\end{align}
where $\Lambda_N \sim 4\pi f$ is the dynamical scale of $SU(N)$ and we determined  the factor of $4\pi$s by the naive dimensional analysis~\cite{Manohar:1983md,Luty:1997fk,Cohen:1997rt}. These terms are the origin of the $S-H$ coupling in Eq.~\eqref{eq:full model}. The relative phase between $y \bar{y}$ and the masses of $L$ and $E$ gives non-zero $\delta$.
Closing the Higgs loop, a potential of $S$ is generated,
\begin{align}
    \frac{\Lambda_N^4} {(16\pi^2)^2} y \bar{y} e^{iS/f} {\rm ln}  \frac{\Lambda_N}{m_{D}}.
\end{align}
In terms of the effective theory in Eq.~\eqref{eq:SH_UVC}, the Higgs loop has a cut-off at $ \Lambda_H \sim m_{D} \sim g f$.
For the parameter region with $m_S = O(10)$ GeV where $f=O(1)$ TeV,
the naturalness condition in Eq.~\eqref{eq:naturalness final} is satisfied.

We next discuss a model where $SW\tilde{W}$ coupling is suppressed beyond $\alpha_2/(8\pi f)$, so that the observed BAU can be explained if overproduced for $c_W = 1$.
We introduce an $SU(N)$ gauge interaction ($N\geq 3$) with fermions shown in Table~\ref{tab:composite-fermion}. $SU(N)$ has 4 flavors, and the $SU(4)_L\times SU(4)_R$ global symmetry is broken down into $SU(4)_V$, yielding the following 15 NGBs:
\begin{align}
    T(3,0),~ D_1(2,1/2),~D_2(2,1/2),~S_c(0,1),~S'(1,0), S(1,0)
\end{align}
There are two neutral NGBs $S'$ and $S$, and the corresponding spontaneously broken global $U(1)'\times U(1)$ charges of the fermions are also shown in Table~\ref{tab:composite-fermion}. $U(1)'$ has $SU(2)_L\times U(1)_Y$ anomaly, while $U(1)$ does not, so $S$ does not have anomalous couplings to the electroweak gauge bosons.

\begin{table}[!t]
    \centering
    \begin{tabular}{|c|c|c|c|c|}
        \hline
                  & $L$       & $E$       & $\bar{L}$ & $\bar{E}$ \\ \hline
        $SU(N)$   & $N$       & $N$       & $\bar{N}$ & $\bar{N}$ \\
        $SU(2)_L$ & ${\bf 2}$ & ${\bf 1}$ & ${\bf 2}$ & ${\bf 1}$ \\
        $U(1)_Y$  & $-1/4$    & $1/4$     & $1/4$     & $-1/4$    \\ \hline
        $U(1)$    & $1$       & $-2$      & $1$       & $-2$      \\ \hline
    \end{tabular}
    \caption{Fermion contents for a composite UV completion with unsuppressed $SW\tilde{W}$ coupling.}
    \label{tab:composite-fermion2}
\end{table}

\begin{table}[!t]
    \centering
    \begin{tabular}{|c|c|c|c|c|c|c|}
        \hline
                  & $L$       & $E_+$     & $E_-$     & $\bar{L}$ & $\bar{E}_+$ & $\bar{E}_-$ \\ \hline
        $SU(N)$   & $N$       & $N$       & $N$       & $\bar{N}$ & $\bar{N}$   & $\bar{N}$   \\
        $SU(2)_L$ & ${\bf 2}$ & ${\bf 1}$ & ${\bf 1}$ & ${\bf 2}$ & ${\bf 1}$   & ${\bf 1}$   \\
        $U(1)_Y$  & $0$       & $1/2$     & $-1/2$    & $0$       & $-1/2$      & $1/2$       \\ \hline
        $U(1)$    & $0$       & $1$       & $-1$      & $0$       & $1$         & $-1$        \\
        $U(1)'$   & $1$       & $-1$      & $-1$      & $1$       & $-1$        & $-1$        \\ \hline
    \end{tabular}
    \caption{Fermion contents for a composite UV completion with suppressed $SW\tilde{W}$ coupling.}
    \label{tab:composite-fermion}
\end{table}

Non-zero masses of $S'$ and $S$ are given by explicit $U(1)'\times U(1)$ breaking. A mass of $L\bar{L}$ only breaks $U(1)'$, and gives a mass only to $S'$. Nonzero masses of $E_+\bar{E}_+$ and $E_-\bar{E}_-$ break both $U(1)'$ and $U(1)$, so give a mass to $S$ as well as $S-S'$ mixing.
Assuming $m_L \gg m_{E_\pm}$, we may achieve a hierarchy $m_{S'}^2 \gg m_S^2$ as well as a small $S-S'$ mixing. The small mixing introduces a small $SW\tilde{W}$ coupling required for the successful electroweak baryogenesis.

We introduce the following couplings with the Higgs;
\begin{align}
    \label{eq:yukawa2}
    {\cal L}= y_1 H L \bar{E}_+ + \bar{y}_1 H^\dag \bar{L} E_+ + y_2 H^\dag L \bar{E}_- + \bar{y}_2 H \bar{L} E_-.
\end{align}
These couplings preserve $U(1)'$ and $S'$ does not couple to the Higgs. On the other hand, these couplings violate $U(1)$ and generate $H-D_{1,2}$ mass mixing whose phase depends on $S$. The exchange of $D_{1,2}$ generates the couplings of $S$ with the Higgs,
\begin{align}
    \frac{\Lambda_N^4}{16\pi^2} e^{i S/f} |H|^2 \left(
    \frac{y_1 \bar{y}_1}{ m_{D_1}^2} +  \frac{y_2 \bar{y}_2}{ m_{D_2}^2}  \right)  + {\rm h.c.},
\end{align}
The relative phase between $y \bar{y}$ and the masses of $E$ gives non-zero $\delta$.
Closing the Higgs loop, a potential of $S$ is generated,
\begin{align}
    \frac{\Lambda_N^4}{(16\pi^2)^2}e^{iS/f} \left( y_1 \bar{y}_1{\rm ln}  \frac{\Lambda_N}{m_{D_1}} +  y_2 \bar{y}_2{\rm ln}  \frac{\Lambda_N}{m_{D_2}} \right).
\end{align}
Again, the effective cutoff of the Higgs loop is
$ \Lambda_H \sim m_{D_{1,2}} \sim g f$.

Let us comment on the stability of composite particles in the two models described above. All of the mesons are unstable. The lightest $SU(N)$ baryon is stable. If $N$ is even, the lightest one is electroweak neutral and is a good dark matter candidate. With $f = $ few TeV, the mass of the baryon is expected by around several 10 TeV. The freeze-out of the annihilation of the baryon may explain the observed dark matter density.

In the two models described above, the $SU(N)$ gauge theories have three and four light flavors respectively. It is considered that the phase transition of such $SU(N)$ is of first order if $N\geq 3$~\cite{Pisarski:1983ms}. With the dynamical scale around $10$ TeV, the phase transition temperature is also around $10$ TeV. The resultant gravitational-wave signal may be observable.

\subsection{Perturbative UV completion}
\label{sec:perturbative}

We consider the following interactions and masses,
\begin{align}
    \label{eq:modelP}
    y P \bar{L}_1 L_2  + \lambda_1 H \bar{N} L_1 + \lambda_2 H^\dag N \bar{L}_2 + m_1 \bar{L}_1 L_1 + m_2 \bar{L}_2 L_2 + m_N \bar{N}N,
\end{align}
where $P$ is a complex scalar, $L_{1,2}$ and $\bar{L}_{1,2}$ are $SU(2)_L$ doublet fermions, and $N$ and $\bar{N}$ are singlet fermions. We assume a wine bottle potential of $P$. The angular direction of $P$ is identified with $S$.
This model is characterized by the collective symmetry breaking~\cite{Arkani-Hamed:2001nha,Arkani-Hamed:2002ikv}, where the shift symmetry of $S$ is violated only if all of $y$, $\lambda_1$,  $\lambda_2$, $m_1$ $m_2$, and $m_N$ are non-zero. For example, when $\lambda_1=0$, there is a $U(1)$ symmetry under which $P(1)$, $L_1(-1)$, $\bar{L}_1(1)$, with other field having vanishing $U(1)$ charges.

Because of the collective symmetry breaking, a one-loop correction to the $S-H$ coupling is finite:
\begin{align}
    \frac{y \lambda_1 \lambda_2}{16\pi^2 }\frac{m_1 m_2 m_N f}{\Lambda^2}e^{iS/f} |H|^2 + {\rm h.c.},~~\Lambda = {\rm max}(y \hvev{P},m_1,m_2,m_N).
\end{align}
A two-loop correction to the potential of $S$, which arises as a tadpole term of $P$, is logarithmically divergent,
\begin{align}
    \frac{y \lambda_1 \lambda_2}{(16\pi^2)^2} m_1 m_2 m_N f {\rm ln} \frac{\Lambda_c}{\Lambda} e^{iS/f}  + {\rm h.c.},
\end{align}
where $\Lambda_c$ is the cutoff of the model in Eq.~\eqref{eq:modelP}.
One can see that $\Lambda_H\sim \Lambda$, which is given by the largest fermion mass scale. Even if $m_S=O(10)$ MeV, where $f= O(10^{6-7})$ GeV is much above the upper bound on $\Lambda_H$  $\sim $ few TeV, the Higgs loop may be cut off at the scale much below $f$, and the naturalness bound can be satisfied.

For $\lambda_{1,2}=0$, the shift symmetry of $S$ is exact as mentioned above, and the shift symmetry does not have an electroweak anomaly. This means that $SW\tilde{W}$ coupling vanishes when $H=0$. Indeed, the determinant of the mass matrix of the electromagnetically charged fermions is independent of $P$, and that of the neutral ones are $m_1m_2m_N + y \lambda_1 \lambda_2 |H|^2P$. The coefficient of $W\tilde{W}$ is proportional to the log of the determinant. The resultant $SW\tilde{W}$ coupling is suppressed by the ratio between $|H|^2$ and the masses of the fermions and is too small to generate the observed BAU.
We may introduce mass terms $\ell \bar{L}_{1,2}$ to generate $\partial S\ell^\dag \bar{\sigma}\ell$ coupling, which can generate the observed BAU.

\section{Experimental signals}
\label{sec:probe}

In this section, we discuss various ways to probe the ALP-assisted model.

\subsection{Higgs exotic decay}

The ALP-Higgs coupling leads to exotic Higgs decay, $h \rightarrow S S$.
As an effective probe of scalar extensions of the SM, the Higgs exotic decay has been intensively searched at the LHC for various SM final states of $S$ with the branching ratio determined from the mixing with the Higgs, see Ref.~\cite{Carena:2022yvx} for the most updated review. High luminosity LHC (HL-LHC) and the Higgs Factory are expected to put a limit on the exotic decay branching ratio by 1-2 orders of magnitude stronger than current searches.
In this model, the relevant operators up to dim-4 are $\lambda \sqrt{2} v h^3$, $A S h^2$, and $A S^2 h \sqrt{2}v/(2f)$.
The $hSS$ coupling after the singlet-Higgs mixing is
\begin{align}
    \label{eq:higgs decay coupling}
    g_{hSS} = 3 \lambda \sqrt{2} v \sin^2 \theta \cos \theta + A \sin \delta \left( \frac{3}{2}\sin^3 \theta - \sin \theta \cos^2 \theta \right) + \frac{A}{f} \sqrt{2} v \cos \delta \left( \frac{1}{2}\cos^3 \theta - \sin^2 \theta \cos \theta \right).
\end{align}
The exotic decay rate is
\begin{align}
    \label{eq:higgs decay br}
    \Gamma_{hSS} = \frac{g_{hSS}^2}{32 \pi m_h} \sqrt{1 - 4 \frac{m_S^2}{m_h^2}}.
\end{align}

The effective $g_{hSS}$ coupling has a term $A v f^{-1} \cos \delta \cos^3 \theta$ where $\sin \theta$ does not appear directly. Small mixing angle $\theta$ thus suppresses this term only by making $A$ smaller. This suppression is linear, leading to a slowly decreasing exotic branching ratio for decreasing $\sin\theta$.
Though the branching ratio is beyond the reach of the current existing limit, the future Higgs exotic decay search at the Higgs factory (and HL-LHC for some specific parameter region) can probe this model further. In Fig.~\ref{fig:final result}, we show the future projection for HL-LHC and Higgs factory for different values of $c$ and $\delta$. The current existing limit curves are outside the plot range.

\subsection{Scalar direct production}
The mixing between $S$ and $h$ induces interaction between $S$ and SM gauge bosons. The vertex $SZZ$ is used to search for $S$ at the LEP. Searches are performed independently of the decay modes of $S$ up to $15~\gev$ by the L3 Collaboration~\cite{L3:1996ome} and up to $100~\gev$ by the OPAL Collaboration~\cite{OPAL:2002ifx}. Searches assuming $S$ decaying into $b \bar{b}$ or $\tau \bar{\tau}$ is also performed up to $100~\gev$~\cite{LEPWorkingGroupforHiggsbosonsearches:2003ing}.
In this paper, we assume that $S$ does not decay into the dark sector and choose the most stringent bound mentioned above for each mass, leading to the purple-shaded region in Fig.~\ref{fig:final result}. The most stringent bound comes from the decay-independent search for $m_S \lesssim 17~\gev$ while from the $b \bar{b}$ final state search for larger masses. The latter can be relaxed if $S$ dominantly decays into the dark sector.

\subsection{Rare meson decay}
The mixing also leads to extra decay channels of mesons such as $B$ and $K$ mesons. For a comprehensive review, see~\cite{Beacham:2019nyx}. For the most recent updated review for experimental searches, see~\cite{Antel:2023hkf}. Here we summarize the relevant searches; the current limit and projected searches are shown in Fig.~\ref{fig:final result} by the shaded region and dashed line, respectively.

The $B$ meson can decay into $K+S$ with $S$ decaying into a muon pair. This decay chain is searched at the LHCb experiment for $200~\mathrm{MeV} \leq m_S \leq 4~\gev$~\cite{LHCb:2015nkv,LHCb:2016awg}, constraining the mixing angle to be smaller than $10^{-3}$, as shown in the upper panels in Fig.~\ref{fig:final result}.
The bound is avoided if $S$ dominantly decays into dark-sector particles~\cite{Das:2009ue}.

If $S$ is lighter than the $K$ meson, it leads to an extra decay channel $K \rightarrow \pi S$ with $S$ further decaying into SM particles (most likely electron or muon pair if mass allowed).
If $m_S$ is smaller than $2m_\mu$, the decay rate is small and $S$ is long-lived at the experimental scale and can thus be regarded as a stable particle, invisible final state.
The NA62 experiment performed searches for long-lived, invisible $S$ via charged kaons~\cite{NA62:2021zjw} and neutral kaons~\cite{NA62:2020pwi}, respectively. E949 experiment searched for charged-kaon decay $K^+ \rightarrow \pi^+ + \text{invisible}$~\cite{E949:2004uaj,BNL-E949:2009dza}, which was reinterpreted as the bound for dark decay channel for ALPs in Ref.~\cite{Dolan:2014ska}.
These experiments provide strong constraints on the parameter region for this model.

Most of the currently allowed parameter space in the MeV scale can be probed by future beam-dump experiments. NA62 Run 2 is expected to finish in 2025~\cite{Antel:2023hkf}. The proposed High Intensity Kaon Experiments (HIKE) project (also known as KLEVER), utilizing a CERN beam, is expected to perform searches at a 15-year time scale.

\subsection{Heavy particles at colliders}

The UV completion of the model generically predicts new particles beyond $S$.
We expect new particles with a mass scale $f$ in UV models for the ALP. We show the contours of $f=1$ TeV by pink dashed lines in Fig~\ref{fig:final result}.
For the GeV scale $S$, the contours show up in the viable parameter regions
and to the right of these lines, $f < 1~\tev$ and there may be observable collider signals from new heavy particles.
For the MeV scale $S$, on the other hand, $f$ varies from $10^6$ GeV to $10^7$ GeV and we do not expect signals from new particles associated with $f$.

We also expect particles associated with the cutoff $\Lambda_H = O(1-10)$ TeV.
In the composite/perturbative UV completion, electroweak charged NGBs/fermions with a mass around $\Lambda_H$ are predicted.

\subsection{Electron electric dipole moment}
\label{sec:eEDM}

\begin{figure}[!t]
    \centering
    \includegraphics[width=0.325\linewidth]{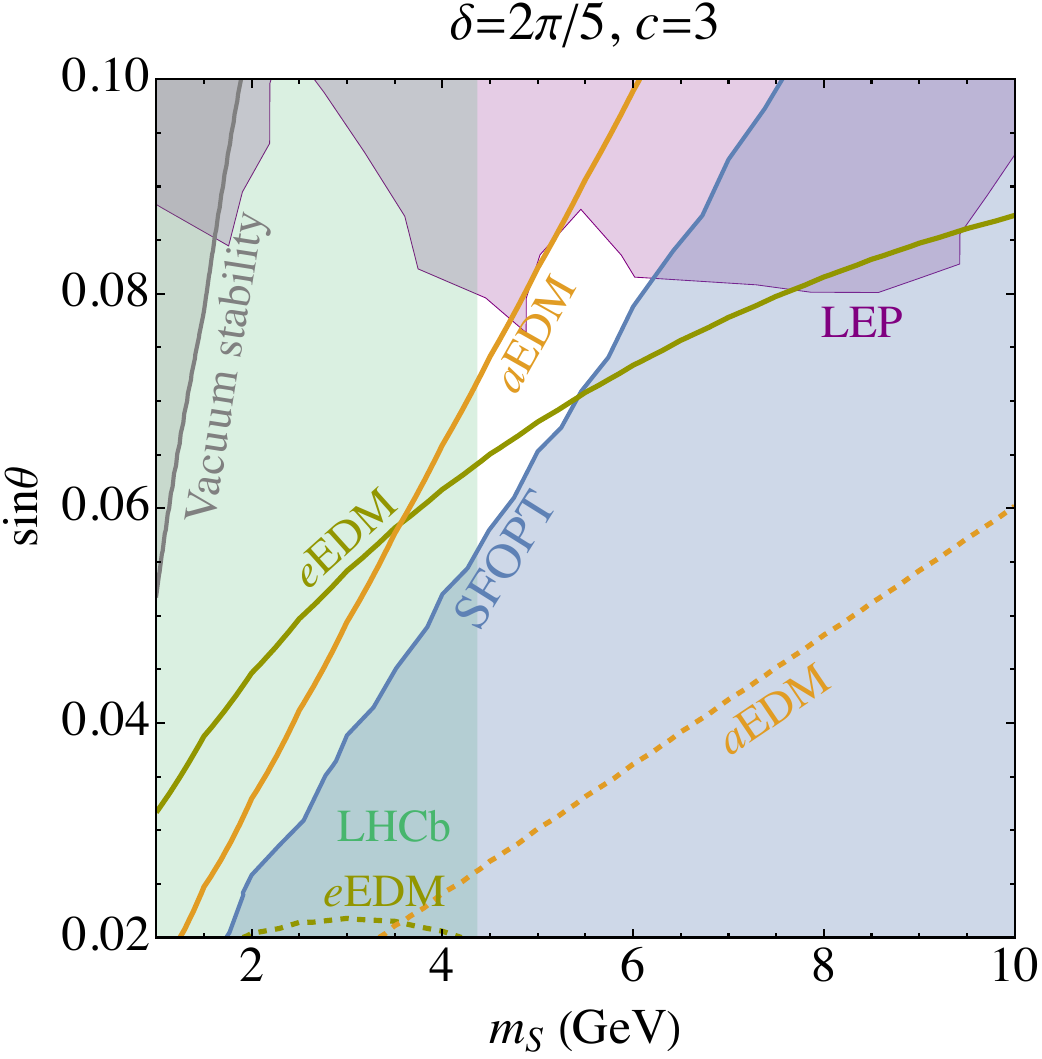}
    \includegraphics[width=0.325\linewidth]{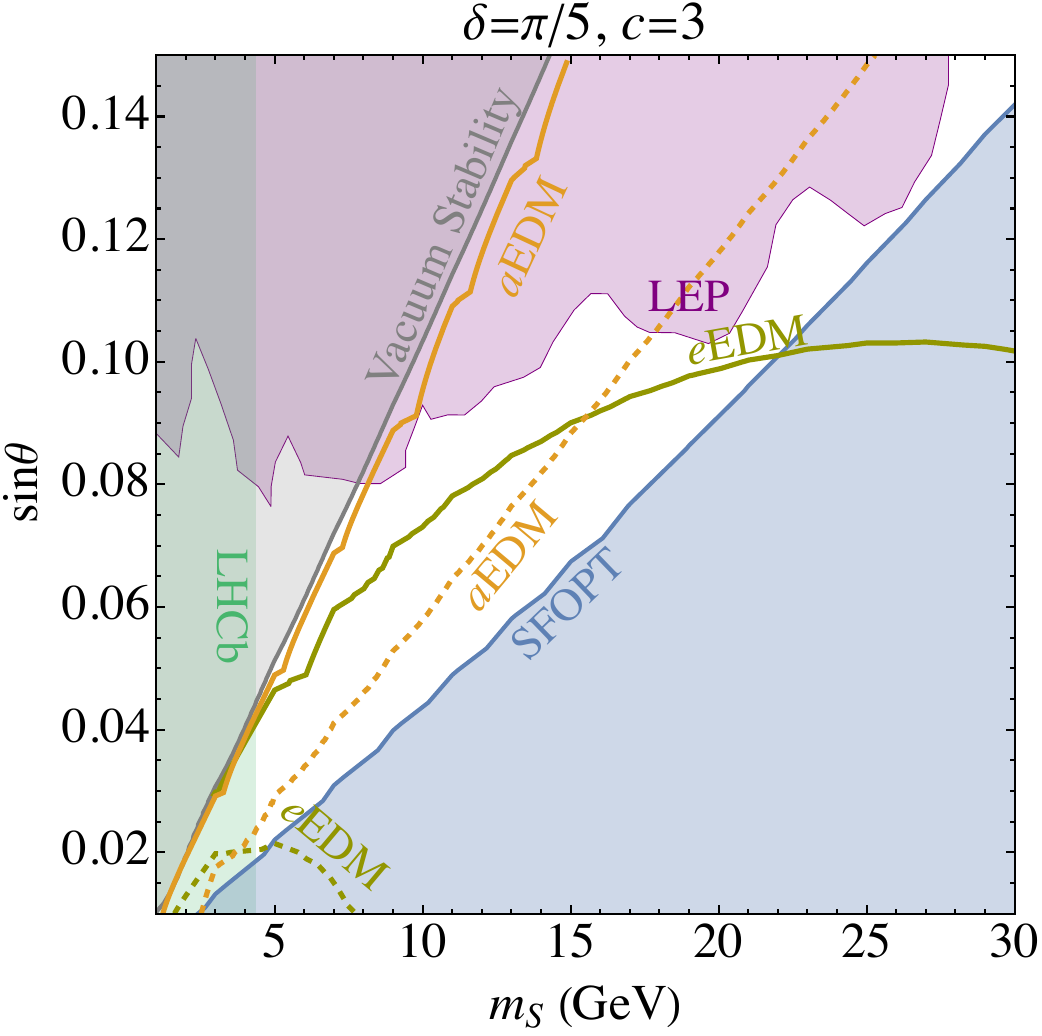}
    \includegraphics[width=0.325\linewidth]{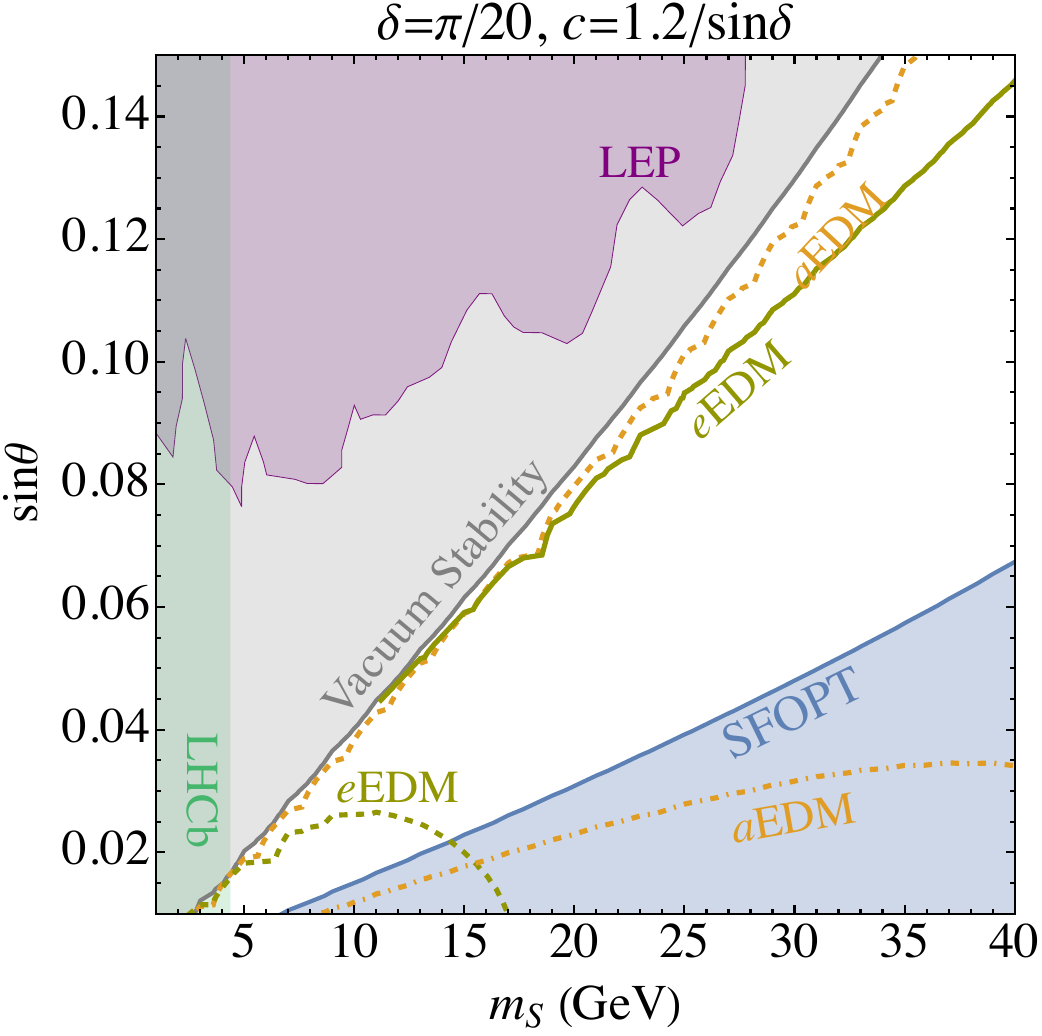}

    \vspace{0.3cm}

    \includegraphics[width=0.325\linewidth]{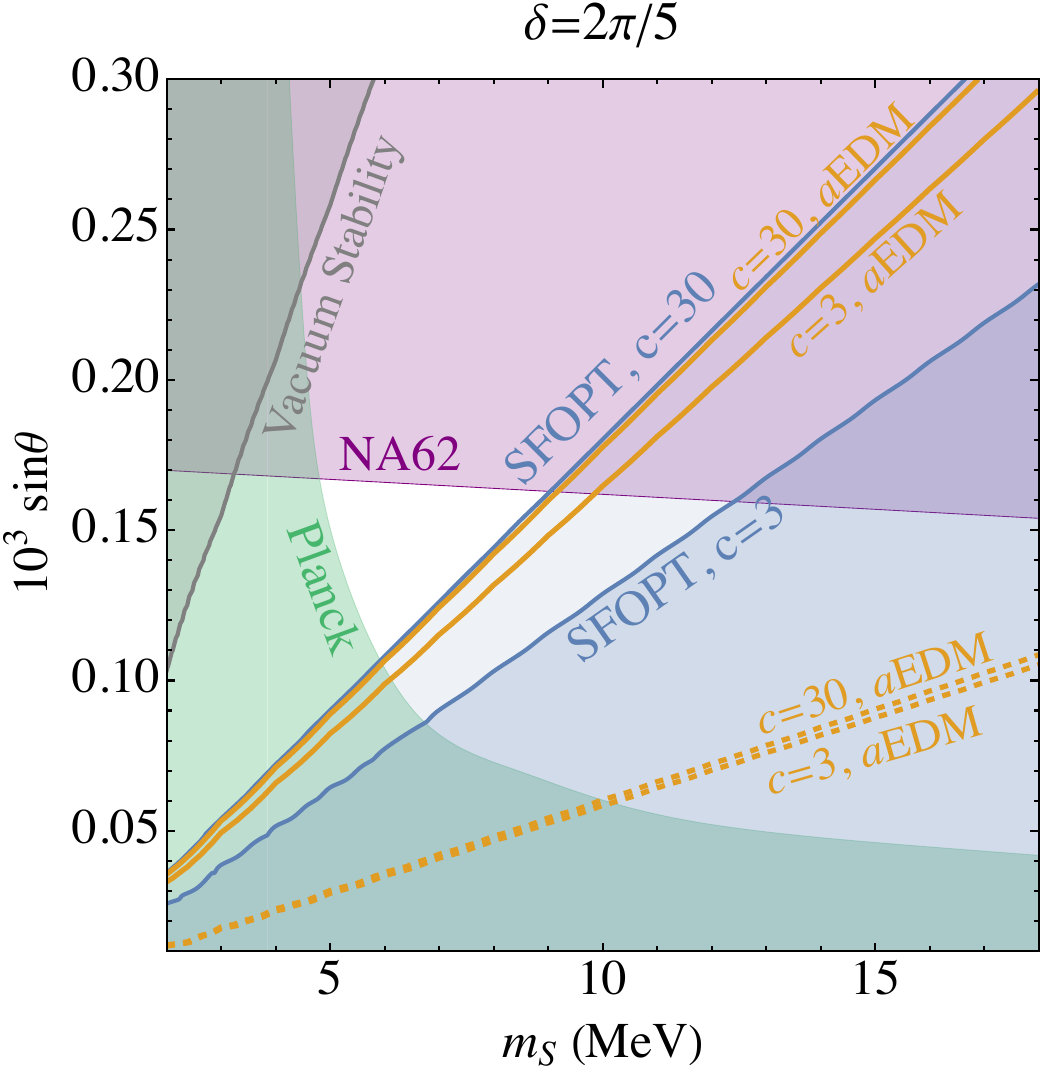}
    \includegraphics[width=0.33\linewidth]{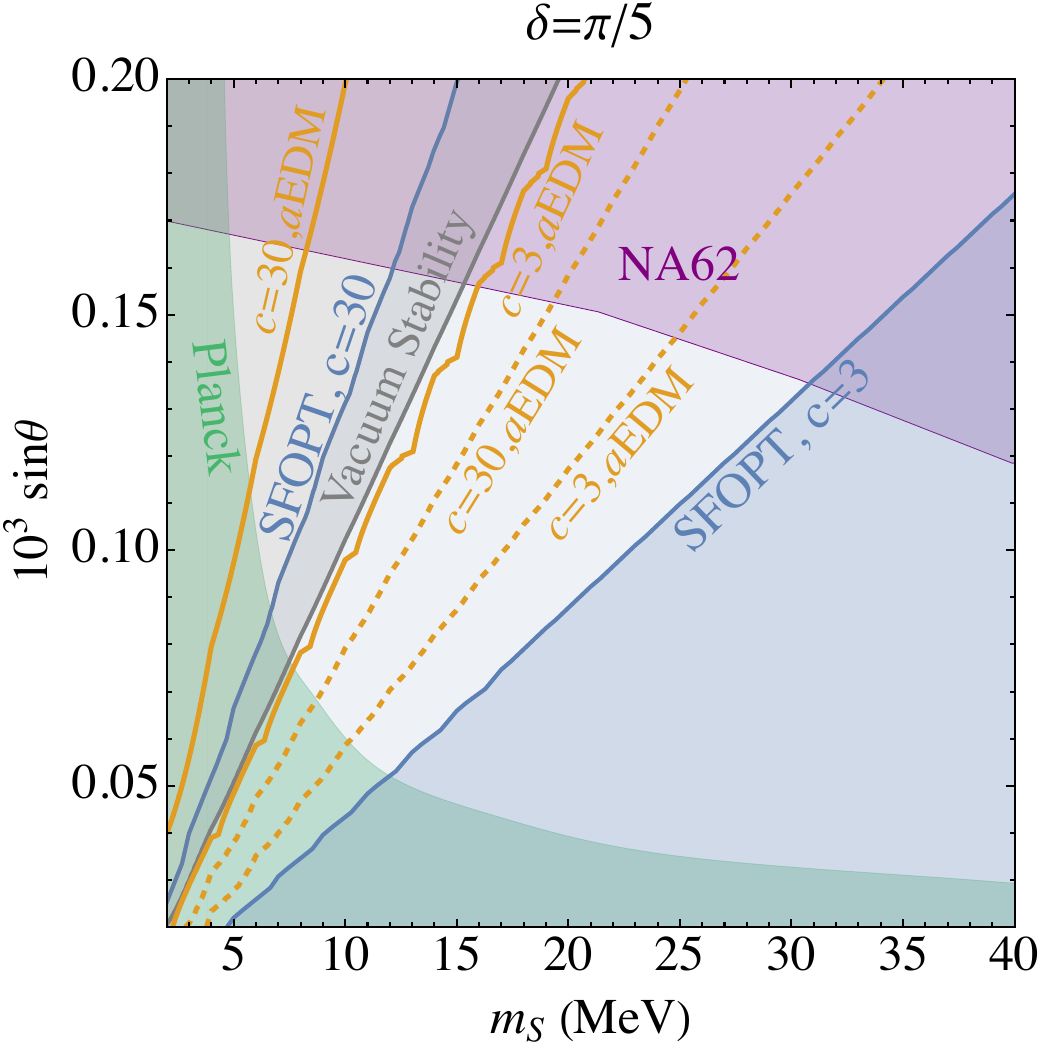}
    \includegraphics[width=0.325\linewidth]{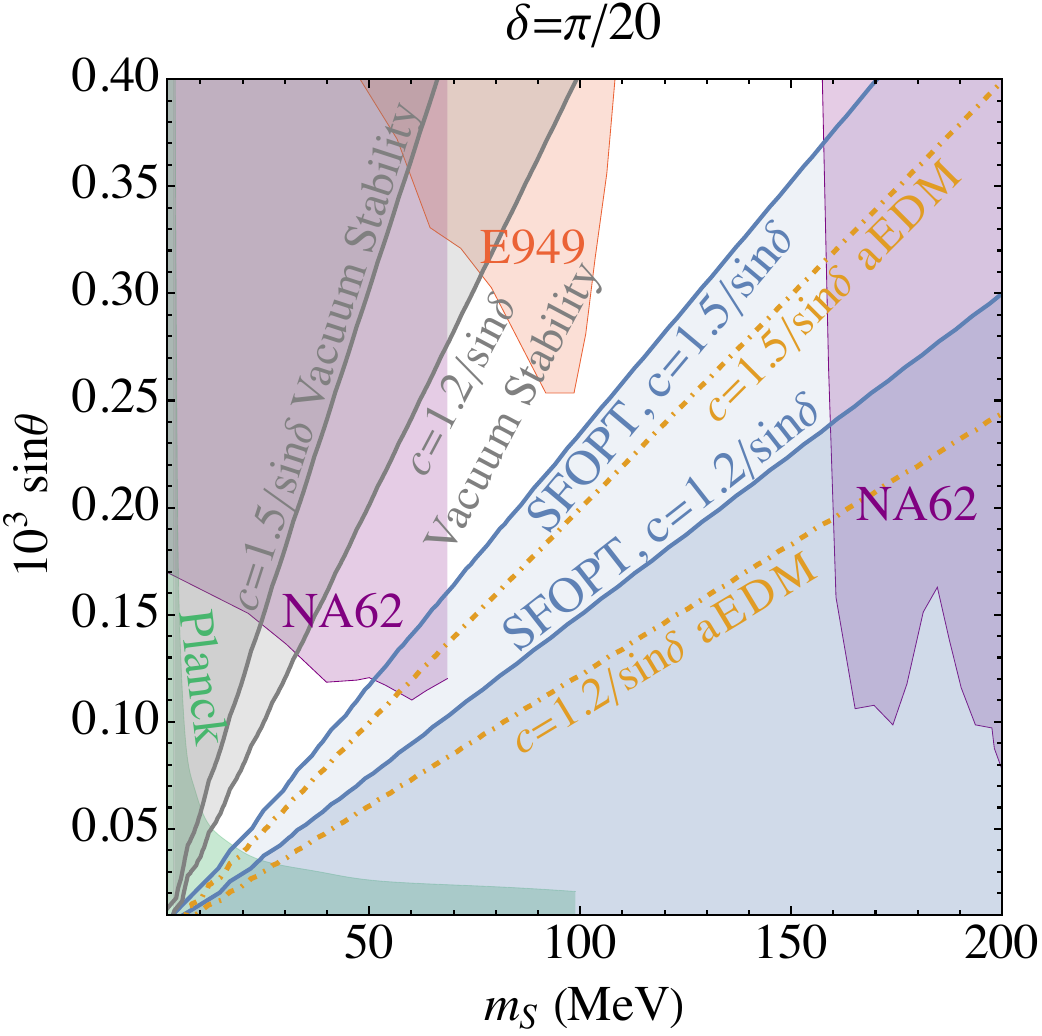}
    \caption{The EDM constraints and future prospects. The regions above the curves are excluded or will be probed.
        Dark green curve: electron EDM discussed in Sec.~\ref{sec:eEDM}.
        Orange curve: atomic EDM discussed in Sec.~\ref{sec:nEDM}.
        For GeV scale $m_S$, we only show one value of $c$ for better visibility of the plots.
        Current EDM contraints are shown by solid curves while the future prospect is plotted by dashed lines for an improvement by a factor of 10 and by dot-dashed lines for a factor of 100.
        The improvement of the EDM constraint will fully cover the plotted range for all the panels.}
    \label{fig:edm-final}
\end{figure}

The electron EDM can be generated via a photon loop and has been computed in Refs.~\cite{Marciano:2016yhf,DiLuzio:2020oah,DiLuzio:2023lmd}.%
\footnote{A $Z$ boson loop also contributes to the EDM. Although the products of the coupling constants in the photon and $Z$ loops are similar, the latter does not have log-enhancement and is expected to be subdominant.}
Here, we consider the $S W \tilde{W}$ coupling in Eq.~\eqref{eq:dim-5 operators} as an example.
By matching our model parameters to Refs.~\cite{Marciano:2016yhf,DiLuzio:2020oah,DiLuzio:2023lmd}, we find
\begin{align}
    \frac{d_e}{e} = \frac{1}{64 \sqrt{2} \pi^4} \frac{c_W e^2 y_e \sin \theta}{f} \ln \left(\frac{m_W}{m_S} \right)~\rm.
\end{align}
We choose $c_W$ so that the observed BAU is explained by local electroweak baryogenesis.
We show the current constraint~\cite{ACME:2018yjb} in Fig.~\ref{fig:edm-final} with dark green solid curves, above which is excluded.
For the following reasons, we only show $c=3$ for $\delta=\pi/5, 2\pi/5$ and $c=1.2/\sin \delta$ for $\delta=\pi/20$:
\begin{itemize}
    \item For $\delta=2\pi/5$, $c=30$ case is already excluded even without the $e$EDM constraint, see Fig.~\ref{fig:final result}.
    \item For $\delta=\pi/5$, $c=30$ curve is very close to the $c=3$ case.
    \item For $\delta=\pi/20$, $c=1.2/\sin \delta$ and $c=1.5/\sin \delta$ have distinguishable $e$EDM curves.
          Their exclusion curves have a similar shape, but are located at different regions.
          Though they prefer different parameter spaces, the total sizes of the allowed parameter space are similar to each other.
          For a better visibility of the plot, we choose to plot only one of them.
\end{itemize}

From the plots, one can see that the electron EDM constraint is not as severe as most of the scalar extension models.
For a small mixing angle, the EDM itself is small. For a large mixing angle, the large filed value shift of $S$ requires $c_W < 1$, and the EDM is suppressed accordingly.
However, most portion of the viable parameter space of the-GeV scale $m_S$  can still be probed in the future if the EDM reach is improved by a factor of $10$~\cite{Vutha:2009ux};
the future prospect is shown in Fig.~\ref{fig:edm-final} by dashed curves.
If the improvement is by a factor of $100$, then all the parameter space of the GeV scale $S$ will be probed.
The electron EDM for the MeV scale $m_S$, on the other hand, is more suppressed by the small mixing angle and large $f$, and it requires an improvement of 3-4 orders of magnitude to probe the parameter space.

The BAU may be explained by the fermionic coupling $c_\psi \partial_\mu S \psi^\dagger \bar{\sigma}^\mu \psi/f$ instead of the $W$ coupling. There are two competing impacts on the EDM. For $c_{\psi} = c_W$,
1) the fermion loop generates a larger electron EDM than the photon loop, but 2) the BAU created by the fermion coupling is larger than the $W$ coupling; see Eq.~\eqref{eq:baryon number}.
We found that the latter effect wins and the fermion model predicts a slightly smaller EDM than the $W$ model. The qualitative conclusion for the $W$ model holds for the fermion model.

\subsection{Atomic electric dipole moment}
\label{sec:nEDM}
The ALP couples to the nucleons and electrons as a CP-even scalar through the mixing with the Higgs. In the model of local electroweak baryogenesis, the ALP may couple to fermions as a CP-odd scalar. If the latter coupling exists for the electron, up quark, or down quark, the exchange of the ALP between nucleons and electrons induces atomic EDMs.

The scalar coupling is model-independently given by
\begin{align}
    - 0.3 {\rm sin}\theta  \frac{m_N}{\sqrt{2}v} S \bar{N}N -  {\rm sin}\theta  \frac{m_e}{\sqrt{2}v}S\bar{e}e,
\end{align}
where the numerical factor of $0.3$ is determined using the trace anomaly~\cite{Shifman:1978zn,Barbieri:2016zxn}. The pseudo-scalar coupling is
\begin{align}
    \label{eq:Sff}
    -i c_\psi \frac{m_\psi}{f} S \bar{\psi}\gamma_5 \psi.
\end{align}

The up-to-date constraint on the ALP-nucleon coupling from atomic EDM ($a$EDM) is derived in Ref.~\cite{Dzuba:2018anu}.
In Fig.~\ref{fig:edm-final}, the constraint is translated to the mixing angle and shown by solid orange curves, assuming that the coupling in Eq.~\eqref{eq:Sff} exists for the electron. The coupling constant $c_e$ is again adjusted to reproduce the observed baryon asymmetry. Parameter spaces above the curves are excluded.
Similar to the electron EDM, improvement by a factor of $10$ and $100$ is assumed for the dashed and dotted-dashed curves. An improvement of $10$ will cover most of the parameter spaces. The improvement by $100$, on the other hand, will fully cover the parameter space, and the contours are below the plotted range except for the $\delta = \pi/20$ case.
If the coupling in Eq.~\eqref{eq:Sff} exists for the up or down quarks rather than the electron, the atomic EDM will change by an $O(1)$ factor. We note that if the coupling does not exist for the first generation of fermions, the atomic EDM is highly suppressed, but is still generated by loop-generated axion-quark or electron couplings.
For example, if the ALP couples to the $W$ boson, the loop-generated atomic EDM will be smaller by a factor of $10^{-4}$ than the case with a tree-level ALP-first generation fermion couplings. An improvement of 5-6 orders of magnitude in the experiments is required to probe the viable parameter space.

\subsection{Effective neutrino number}
For $m_S$ around a few MeV, $S$ is kept in thermal equilibrium in the early universe as late as when neutrinos decoupled from the thermal bath.
The energy of $S$ is transferred into photons and electrons and thus dilutes the neutrino energy density relative to photons, leading to a negative contribution to the effective neutrino number, $N_{\rm eff}$. The constraint on the parameter space and the future projection are derived in~\cite{Ibe:2021fed} using the Planck 2018 data~\cite{Planck:2018vyg} and the future CMB-S4~\cite{CMB-S4:2016ple}. We show the constraint and projection with the green shaded region and dashed line in Fig.~\ref{fig:final result}, respectively.

\section{Conclusion and discussion}
\label{sec:conclusion}
ALPs, as pNGBs, are naturally light and weakly-interacting. In this paper, we investigated the coupling between an ALP and the Higgs to enhance the strength of EWPT and identified the viable parameter space.
In comparison to the early works on this model, we performed full one-loop effective potential computation, including CW corrections and thermal resummation that can significantly reduce the strength of the EWPT.
We found that the EWPT can still be of strong first order in a wide range of parameter space. The ALP can be at the MeV or GeV scale with the mixing angle with the Higgs $O(10^{-1})$ and $O(10^{-4})$, respectively.
As the displacement of the field value of the ALP becomes closer to the decay constant of the ALP, the required mixing angle to achieve SFOPT becomes smaller.

We investigated the two-field phase transition dynamics. The duration of the phase transition is shorter (i.e., a larger $\beta/H$ parameter) for lighter ALPs. In the viable parameter region, gravitational-wave signals are too weak to be detected.

Various ways to probe this model are discussed. For the GeV scale ALP, scalar direct production at the LEP and rare B meson decay provide stringent constraints. The allowed parameter space can be probed by Higgs exotic decay in future collider experiments. For the MeV scale ALP, existing limits come from rare kaon decay and the CMB observation of the effective neutrino number. Future CMB-S4 observation and rare kaon decay experiments can probe most of the currently allowed parameter space.

Baryon asymmetry can be produced by the coupling of the ALP with the $SU(2)_L$ gauge boson or $SU(2)_L$-charged fermions and the local EWBG mechanism.
The observed baryon asymmetry can be produced without violating the current electron EDM bound. Future experiments can fully probe the GeV scale ALP. The MeV scale ALP can be probed if the ALP has a CP-odd coupling with the electron, up quark, or down quark.

We provided two UV completions of the model by composite dynamics or a perturbative fundamental complex scalar field. The composite one explains the smallness of the decay constant of the ALP and is consistent with the GeV scale ALP. The latter one is consistent with ALPs at both the MeV and GeV scales, but the smallness of the decay constant should be explained by further extensions of the model, such as supersymmetry, which is not necessarily at the TeV scale.

In summary, this model has a naturally light and weakly-interacting scalar that enhances the strength of EWPT, in comparison to those traditional scalar extensions whose extra singlet scalar is typically heavy and strongly interacting.  Gravitational-wave signals are too weak, but instead, this ALP can be probed by the EDM, rare meson decay, CMB observation, and Higgs exotic decay, which opens up a window to probe the strong first-order electroweak phase transition.

\section*{Acknowledgement}
We thank Maxim Pospelov for pointing out the constraints from atomic electric dipole moments and Philipp Schicho for providing valuable and detailed assistance when we initially tried to use the \verb|DRalgo| package.
We thank Peizhi Du, Claudius Krause, and Tong Ou for useful discussions.
K.H.~was partly supported by Grant-in-Aid for Scientific Research from the Ministry of Education, Culture, Sports, Science, and Technology (MEXT), Japan (20H01895) and by World Premier International Research Center Initiative (WPI), MEXT, Japan (Kavli IPMU).
I.R.W was supported by the DOE grant DE-SC0010008.
The \href{https://gitlab.com/claudius-krause/ew_nr}{EW-NR repository on GitLab}, wrote for Ref.~\cite{Carena:2021onl}, is used as a template for Python coding.
The Feynman diagrams are made by the public Tikz-Feynman package~\cite{Ellis:2016jkw}.

\appendix

\section{Phase transition strength in SM with lighter Higgs mass and computational uncertainties}
\label{sec:app}

In the introduction, we reviewed the past literature and concluded that SFOPT is unlikely to be achieved for the simplified model of Eq.~\eqref{eq:origianl 1d} in Refs.~\cite{Das:2009ue,Espinosa:2011ax,Harigaya:2022ptp}.
Except for the negligible extra scalar contribution to the thermal barrier, the potential along the valley is equivalent to the SM with a small quartic coupling, i.e., a light Higgs.
Here, we summarize the numerical result of the PT strength for the SM with a lighter Higgs.

\begin{figure}[!t]
    \centering
    \includegraphics[width=0.85\linewidth]{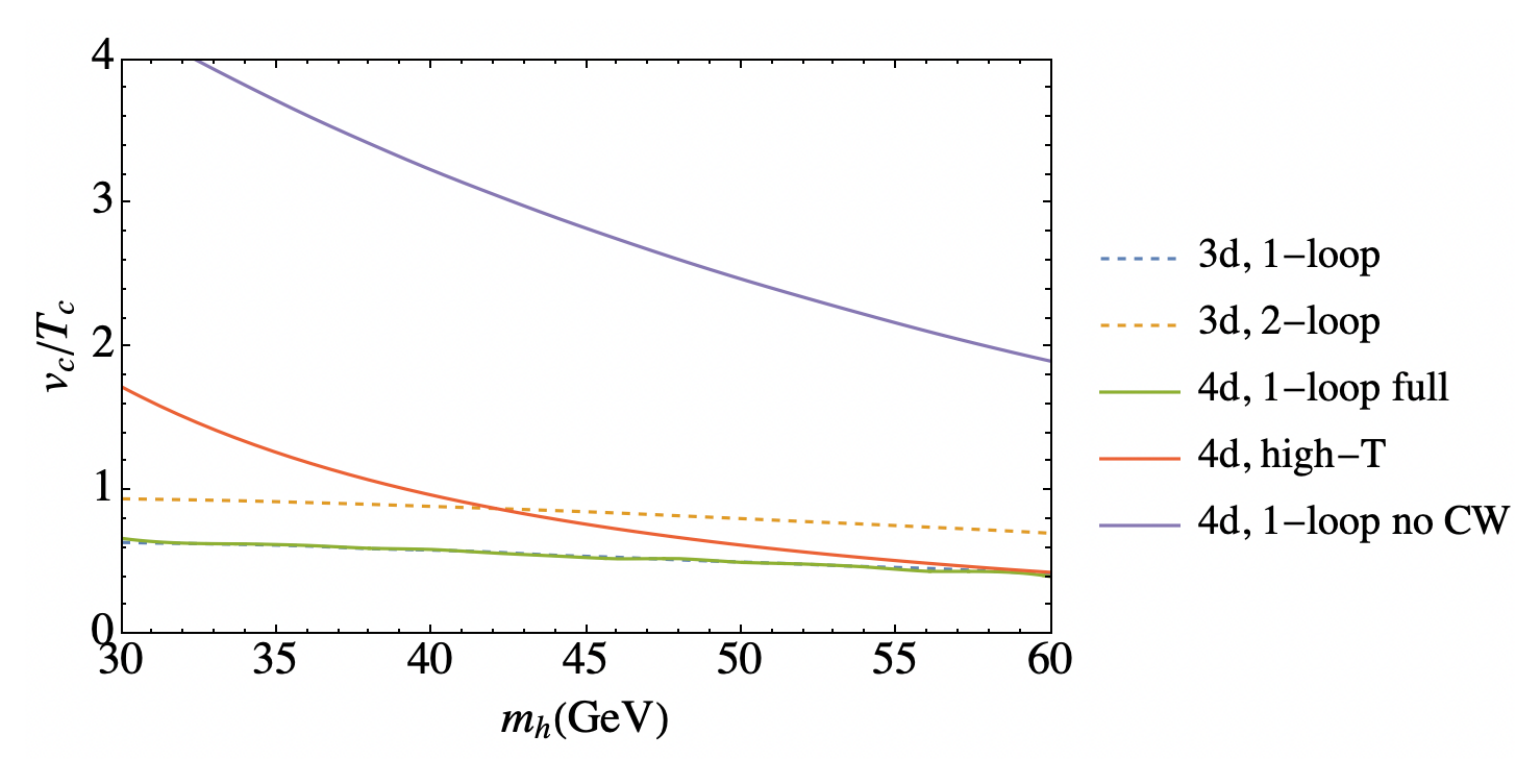}
    \caption{Comparison between different computation methods for the finite-temperature effective potential.}
    \label{fig:VFTcompare}
\end{figure}

Besides the computation we used in the main text, we have the following computation methods:
\begin{itemize}
    \item Dimensional reduction with a two-loop effective potential (3d 2-loop). This computation uses the dimensional reduction method~\cite{Ginsparg:1980ef,Appelquist:1981vg,Braaten:1995cm,Kajantie:1995dw} to integrate out heavy modes and work in a 3-dimensional effective theory. It computes the effective potential up to two-loop. In addition, this computation method uses running coupling to improve the computation at finite temperatures. The input parameter is still $\overline{\rm MS}$ renormalized under the framework of Ref.~\cite{Buttazzo:2013uya}. This is the state-of-the-art computational method to compute the thermal effective potential.
    \item Dimensional reduction with a one-loop effective potential (3d 1-loop).
    \item Traditional one-loop computation, which is used in the main text (4d 1-loop full).
    \item High-temperature expansion, i.e., Eq.~\eqref{eq:highT simplified} in the SM case (4d high-T). To simplify the calculation,
          one widely-used way is to replace the resummation procedure by multiplying the $E_{\rm SM}$ term by a factor of $2/3$ to screen out the longitudinal bosonic modes.
    \item Tree-level zero-temperature potential plus one-loop finite-temperature correction without resummation (4d 1-loop no CW).
\end{itemize}

We compare the result of $v_c/T_c$ in Fig.~\ref{fig:VFTcompare}.
One can see that
the 4d 1-loop no CW computation
overestimates the PT strength by an $O(1)$ factor.
The 4d high-T computation,
although the difference is not that large, also significantly overestimates the PT strength.
The 4d 1-loop full computation
predicts smaller $v_c/T_c$ and agrees with the 3d 1-loop computation. The 3d 2-loop computation predicts stronger PT, but is not yet enough to avoid the wash-out of baryon asymmetry after the EWPT.
Note that the result of the 3d 2-loop computation is different from that of the 3d 1-loop computation by 50\%.
This can be attributed to the cancellation of the thermal mass with the zero-temperature mass around the PT, which makes the one-loop thermal mass plus the zero-temperature mass comparable to the two-loop thermal mass~\cite{Kajantie:1995dw}.

\bibliographystyle{utphys}
\bibliography{ALP_EWPT}

\end{document}